\documentclass{aa}
\usepackage{amssymb,amsmath,times}
\usepackage{mathtools}
\usepackage{lscape}
\usepackage{epstopdf}

\def\gsim{ \lower .75ex \hbox{$\sim$} \llap{\raise .27ex \hbox{$>$}} }                              
\def\mujb{$\mu\mathrm{Jy\,beam}^{-1}$}
\def\nh{$N_{\rm H}$}
\def\lsim{ \lower .75ex\hbox{$\sim$} \llap{\raise .27ex \hbox{$<$}} }
\def\ph{ph cm$^{-2}$ s$^{-1}$}

\usepackage{natbib,graphicx}
\usepackage{txfonts}
\begin{document}

   \title{The 999$^{\rm th}$ \textit{Swift} gamma-ray burst: Some like it thermal}

   \subtitle{A multiwavelength study of GRB 151027A}

   \author{F. Nappo
          \inst{1,2}\fnmsep\thanks{\email{francesco.nappo@brera.inaf.it}},                                       
          A. Pescalli\inst{1,2}, G. Oganesyan\inst{3}, G. Ghirlanda\inst{2}, 
          M. Giroletti\inst{4}, A. Melandri\inst{2}, S. Campana\inst{2}, 
          G. Ghisellini\inst{2}, O.~S. Salafia\inst{5,2}, 
          P. D'Avanzo\inst{2}, 
          M.~G. Bernardini\inst{6}, S. Covino\inst{2}, 
          E. Carretti\inst{7},
          A. Celotti\inst{3},
          V. D'Elia\inst{8,9},
          L. Nava\inst{10},
          E. Palazzi\inst{11},
          S. Poppi\inst{7},
          I. Prandoni\inst{4},
          S. Righini\inst{4},
          A. Rossi\inst{11},
          R. Salvaterra\inst{12},
%          C. Stanghellini\inst{4},
          G. Tagliaferri\inst{2},
          V. Testa\inst{7},
          T. Venturi\inst{4},
          S.~D. Vergani\inst{13}
          }

   \institute{Universit\`a degli Studi dell'Insubria, via Valleggio 11, I--22100 Como, Italy              
         \and
             INAF -- Osservatorio Astronomico di Brera, via Bianchi 46, I--23807 Merate (LC), Italy
             \and
             SISSA, via Bonomea 265, I--34136 Trieste, Italy
             \and
             INAF -- Istituto di Radioastronomia, via Gobetti 101, I--40129, Bologna, Italy
             \and
             Universit\`a degli Studi di Milano--Bicocca, Piazza della Scienza 3, I--20126, Milano, Italy
             \and
             Laboratoire Univers et Particules de Montpellier, Universit\'e Montpellier, CNRS/IN2P3, F--34095 Montpellier, France
             \and
             INAF -- Osservatorio Astronomico di Cagliari, via della Scienza 5, I--09047 Selargius (CA), Italy
             \and
             INAF -- Osservatorio Astronomico di Roma, via Frascati 33, I--00040 Monte Porzio Catone (RM), Italy
             \and
             ASI--Science Data Center, via del Politecnico snc, I--00133 Roma, Italy 
             \and
             Racah Institute of Physics, The Hebrew University of Jerusalem, 91907 Jerusalem, Israel
             \and
             INAF--IASF Bologna, Area della Ricerca CNR, via Gobetti 101, I--40129 Bologna, Italy 
             \and
             INAF--IASF Milano, via E. Bassini 15, I--20133 Milano, Italy
             \and
             GEPI -- Observatoire de Paris, CNRS UMR 8111, Univ. Paris-Diderot, 5 Place Jules Jannsen, F--92190 Meudon, France
             }

   \date{Received ; }

% \abstract{}{}{}{}{} 
% 5 {} token are mandatory
 
\abstract
{We present a multiwavelength study of GRB\,151027A. 
This is the 999th gamma-ray burst detected by the {\it Swift} satellite and it has a densely sampled emission in the X-ray and optical band and has been 
observed and detected in the radio up to 140 days after the prompt. The multiwavelength light curve from 500 seconds to 140 days can be modelled  through a standard forward shock afterglow, but it requires an additional emission component to reproduce the early X-ray and optical emission. We present optical observations performed with the Telescopio Nazionale Galileo (TNG) and the Large Binocular Telescope (LBT) 19.6, 33.9, and 92.3 days after the trigger which show a bump with respect to a standard afterglow flux decay and are interpreted as possibly due to the underlying supernova and host galaxy (at a level of $\sim 0.4\, \mu$Jy in the optical $R$ band, $R_{\rm AB}\sim 25$). Radio observations, performed with the Sardinia Radio Telescope (SRT) and Medicina in single-dish mode and with the European Very Long Baseline Interferometer (VLBI) Network and the Very Long Baseline Array (VLBA), between day 4 and 140 suggest that the burst exploded in an environment characterized by a density profile scaling with the distance from the source (wind profile). A remarkable feature of the prompt emission is the presence of a bright flare 100 s after the trigger, lasting $\sim$70 seconds in the soft X--ray band, which was simultaneously detected from the optical band up to the MeV energy range. By combining  {\it Swift}--BAT/XRT and {\it Fermi}--GBM data, the broadband (0.3--1000 keV) time resolved spectral analysis of the flare reveals the coexistence of a  non-thermal (power law) and thermal blackbody components. The blackbody component contributes up to 35\% of the luminosity in the 0.3--1000 keV band. The $\gamma$-ray emission observed in {\it Swift}--BAT and {\it Fermi}--GBM anticipates and lasts less than the soft X-ray emission as observed by {\it Swift}--XRT, arguing against a Comptonization origin. The blackbody component could either be produced by an outflow becoming transparent or by the collision of a fast shell with a slow, heavy, and optically thick fireball ejected during the quiescent time interval between the initial and later flares of the burst.}


   \titlerunning{GRB\,151027A}
   \authorrunning{F. Nappo et al.}
   \maketitle
%
%________________________________________________________________

\section{Introduction}

%Understanding the central engine and the emission process of Gamma Ray Bursts 
%(GRBs) passes through the analysis and study of their prompt and afterglow emission components. 
The analysis and study of both the prompt and afterglow emission in gamma-ray bursts (GRBs) is required for a 
complete understanding of their central engine and emission processes.
The {\it Fermi} satellite has shown 
 the presence of long-lasting emission extending up to the GeV energy range (e.g. \citealt{Abdo+09,Ackermann+10,Ghirlanda+10,Ghisellini+10,Guiriec+10,Ackermann+13}) and 
 a sometimes complex coexistence of thermal and non-thermal components during the prompt 
phase observed between 8 keV and a few MeV (\citealt{Guiriec+11, Guiriec+13, Ghirlanda+13, Burgess+14}). 
These observations stimulated the debate on the origin of the prompt emission in GRBs.  
The {\it Swift} satellite has been enriching the observational picture of the afterglow emission 
either directly, by systematic monitoring of the X--ray (0.3-10 keV) light curve from a few tens 
of seconds to months after the trigger (see e.g. \citealt{Gehrels+09}), or indirectly, by triggering 
ground based follow up programs/telescopes in the optical band. 
Still there are several open issues related to the progenitor (both in long and short GRBs), 
regarding the nature of the outflow (magnetic or matter dominated), the emission 
process of the prompt phase, and the circumburst density. 
From the observational point of view it is hard to answer these questions with a few observations per bursts. 
Either statistical studies of well-defined GRB samples (\citealt{Salvaterra+12}; 
\citealt{Hjorth+12}; \citealt{Perley+16}) or single-event modelling like  GRB\,130427A (\citealt{Maselli+14, Vestrand+14, VanDerHorst+14, Perley+14, Bernardini+14, Ackermann+14, Panaitescu+13, Kouveliotou+13, Laskar+13}) seem to be the best approaches to compare theory and observations. 
However, the latter case is possible only in a handful of bursts and still the wealth of 
information (as for GRB\,130427A) does not completely break some parameter 
degeneracies. 
%{\bf METTERE QUALCHE COSA IN PIU SUL PROBLEMA DELLA DENSITA DI 130427A}. 
%Still it is worth 
Nevertheless, it is still important 
to study in detail any new single event which presents peculiar properties 
of either the prompt and/or afterglow emission, especially if with good data quality and coverage. 

GRB\,151027A, the 999th burst detected by the {\it Swift} satellite, is a long bright event 
lasting about 130
%more than 150 
seconds which was followed in the X--ray 
%(by the {\it Swift} autonomous slewing) 
and in the optical and radio bands until five months 
after the burst. 
The event presents unique properties in the prompt emission due to the presence of 
a bright flare (see e.g. \citealt{Burrows+05a, Chincarini+10, Margutti+10, Bernardini+11}), which has been observed from 0.3 keV to $>$MeV (by {\it Swift}/XRT 
and {\it Swift}/BAT and by {\it Fermi}/GBM). 
Here we present the time resolved spectral analysis of the entire prompt emission with 
particular emphasis on the flare, which shows the presence of two independent spectral 
components: a %thermal
 blackbody and a non-thermal cutoff power law. 
We also present the multiwavelength light curve (obtained by combining public and proprietary optical 
and radio observations) and model the emission with a standard afterglow forward shock scenario. 

In \S 2 we describe the multiwavelength data collected in this paper. 
Results of the spectral and temporal analysis of the broadband emission of GRB\,151027A are presented in \S 3, while the modelling of the prompt and 
afterglow emission are presented in  \S 4. 
In \S 5 we discuss our results. 
Throughout the paper a standard flat cosmological model with $H_0= 67$ km s$^{-1}$Mpc$^{-1}$, 
$\Omega_\Lambda=0.7$, $\Omega_{\rm m}=0.3$ is adopted. Errors are given at a 1$\sigma$
confidence level unless otherwise stated.
  
%  Its emission was promptly followed in the X-ray  has a prompt lightcurve characterized by 3 peaks: the first two within ... sec and the last one happening at .. and lasting ... . The last peak has been clearly observed by {\it Fermi}--GBM in $8-1000$ keV range and by {\it Swift}--BAT in the overlapping 15-150 keV energy range. Remarkably, this peak has been detected by XRT in the softer 0.3-10 keV. This is the first burst with a flare detected up to 1 MeV and bright enough to allow broadband (0.3--1000 keV) time resolved spectral analysis. [CITAZIONE A LAVORO TROJA - SU UL GBM/LAT..]  Through the time resolved spectral analysis of the prompt phase we find a general softening of a non--thermal  nction 
%with $\alpha=-1.41 \pm 0.04$ and $E_{\rm peak,obs}=340 \pm 63$ keV.
%The fluence in $10-1000$ keV band is $\mathcal{F}=(1.94 \pm 0.05)\times 10^{-5}$ erg/cm$^2$ that
%corresponds to an isotropic emitted energy $E_{\rm iso}=3.98 \times 10^{52}$ erg.
%Intensive multiwavelength follow up (X--ray, optical and radio) campaign started soon after the optical counterpart detection and provided flux density measurements as early as 20 s, thus overlapping with the prompt high energy emission. 

\section{Multiwavelength data}

In the following section we present both the data sets collected from the literature and our dedicated observations. The reduction and the analysis of our data is described as well.
%In the following section we describe the data sets collected from the literature and the dedicated observations and their analysis that we performed in the different bands.

\subsection{Gamma-ray and X-ray data}\label{HEdata}

\begin{figure*}[!hbt]
   \centering
%   \vskip -4truecm
 % \includegraphics[width=160mm]{IeIIpicco_new_bis_2.eps}
   \includegraphics[width=160mm]{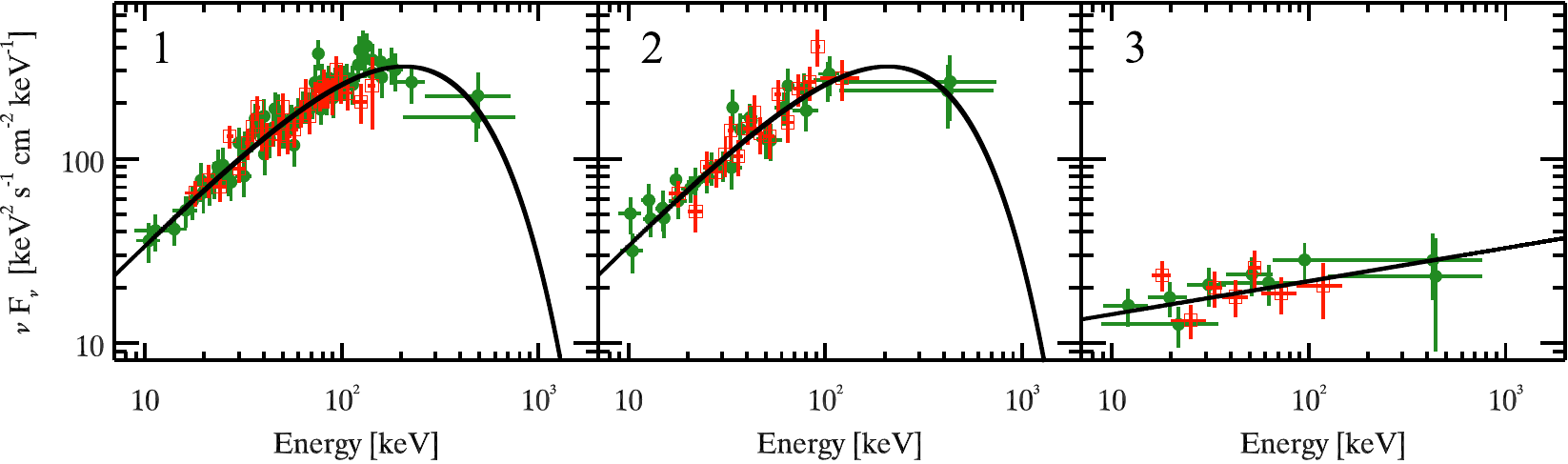}
   \vskip 0.2truecm
   \includegraphics[width=170mm]{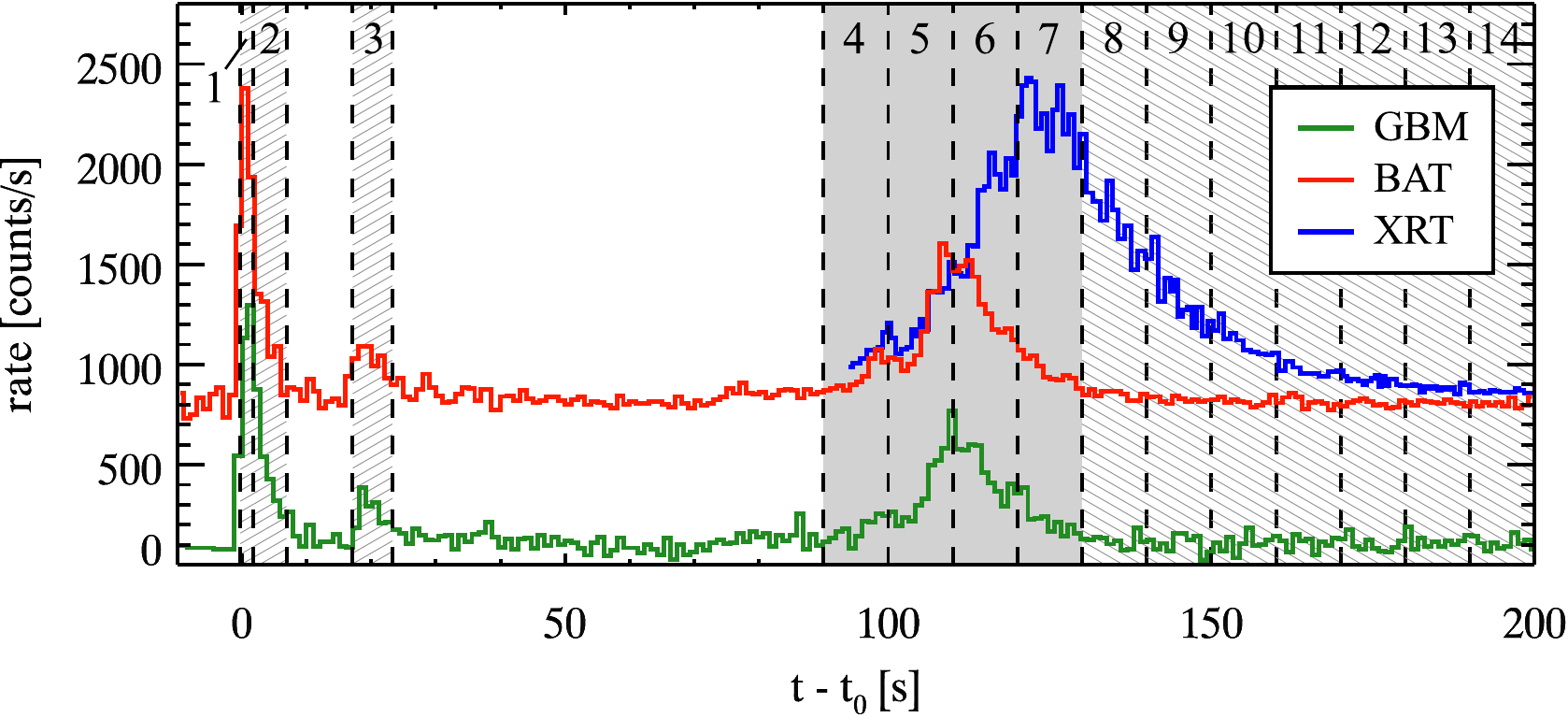}  
   \vskip 0.2truecm	
    \includegraphics[width=160mm]{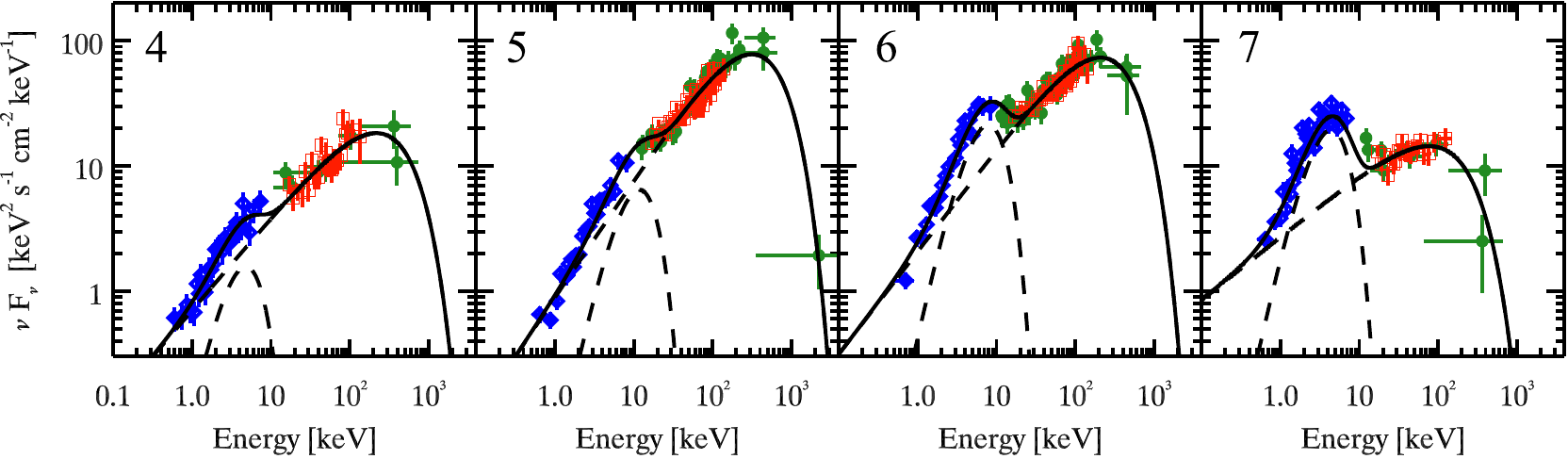}
%   \vskip -2.5truecm
\caption{GRB 151027A rate light curve (middle panel). 
The {\it Fermi}/GBM NaI(0) data (8 keV -- 1 MeV) are shown by the green line. 
The {\it Swift}/BAT (15--150 keV) light curve is shown by the red line. 
The rescaled {\it Swift}/XRT (0.3--10 keV) light curve is shown by the blue line. 
Vertical dashed lines represent the start and stop times of the intervals selected 
for the spectral analysis. 
%{\bf Aggiungere riga verticale aggiungere numeri o colorare per identificare gli spettri ... 
%aggiungere legenda e cambiare i colori.} 
{\it Top panels:} $\nu F_{\nu}$ spectra of the rise (left, \#1) and decay (middle, \#2) phase of the first 
peak and of the second peak (right, \#3), corresponding to the first three hatched regions in the middle panel. 
Green and red symbols show the GBM and BAT data, respectively, and the solid line is the best 
fit model (see Table \ref{tab:prompt}). 
{\it Bottom panels:} $\nu F_{\nu}$ spectra of the third peak corresponding to the shaded 
regions of the light curve (middle panel, \#4-\#7). 
The blue symbols show the XRT spectra. 
The two model components (cutoff power law and blackbody) are shown by the dashed lines and 
their sum by the solid line.  
}
\label{fig:lcprompt}%
\end{figure*}

GRB\,151027A (Maselli et al. 2015) %, GCN 18478) 
was detected and located at 03:58:24 UT by 
the {\it Swift} Burst Alert Telescope (BAT; \citealt{Barthelmy+05}). 
The {\it Swift} X--Ray Telescope (XRT; \citealt{Burrows+05b})  and the Ultra Violet Optical Telescope (UVOT; \citealt{Roming+05}) started 
acquiring data  87 s and 95 s post trigger, respectively, and detected a bright X--ray and optical transient.  
The XRT light curve (limited to the first 200 s since the trigger) is shown in Fig. \ref{fig:lcprompt} 
(blue line) while the full time light curve is shown in Fig. \ref{fig:lcaft}. 
The 15--350 keV energy band BAT light curve has a duration of $T_{90}=130\pm 6$ s (%15--350 keV -- 
Palmer et al. 2015) %, GCN 18496)
 with two main emission episodes (the first  composed of two peaks) 
separated by a quiescent phase of $\sim$ 80 s (see Fig. \ref{fig:lcprompt} -- red line). 
The 15--150 keV band peak flux (corresponding to the first peak at 0.2 s) is 6.8$\pm$0.6 \ph\
and the fluence (7.8$\pm$0.2)$\times10^{-6}$ erg cm$^{-2}$. 

%The preliminary analysis of the time integrated BAT spectrum is consistent with a 
%power law $N(E)\propto E^{-\Gamma}$ with photon index $\Gamma$=1.72$\pm$0.05 (Palmer et al. 2015).

The burst was also detected by the Gamma Burst Monitor (GBM; \citealt{Meegan+09}) on board the {\it Fermi} satellite 
(%trigger number 467611108 / 151027166, 
Toelge et al. 2015) %, GCN 18492) 
and by %the 
Konus--Wind (Golenetskii et al. 2015). The %, GCN 18516). 
{\it Swift}/BAT, {\it Fermi}/GBM (red and green line in the middle panel of Fig. \ref{fig:lcprompt}, respectively), 
and Konus--Wind light curves show similar temporal properties. 
%\footnote{GRB 151027A was also 
%detected by the INTEGRAL SPI (Toelge et al. 2015).}. 
The wide energy ranges of the GBM (8 keV -- 1 MeV) and Konus--Wind (20 keV -- 5 MeV) show that 
the time-averaged spectrum is best fit by a cutoff power law model 
%$N(E)\propto E^{-\Gamma} \exp[E(\Gamma+2)/E_{\rm peak}]$ 
with $\Gamma=1.41 \pm 0.04$
and 
%the cutoff energy 
$E_{\rm peak}=340\pm63$ keV (GBM -- Toelge et al. 
2015)\footnote{The Konus--Wind spectrum, with respect to the GBM, has 
an identical $\Gamma$ but a somewhat smaller $E_{\rm peak}=173^{+135}_{-46}$ keV (Golenetskii et al. 2015). }. 
The GRB fluence in the 10 keV -- 1 MeV energy range, as measured by the GBM spectrum,  is 
(1.94$\pm$0.09)$\times 10^{-5}$ erg cm$^{-2}$ and the photon peak flux 11.37$\pm$0.34 \ph. 

%------------------------------------ FIGURA ----------------------------------------------
%-------------------------------------------------------------------------------------------------

%Early XRT data while the source was still bright allowed for an estimate of the source   
%$z=0.38\pm$0.17 and an intrinsic column density \nh=(3.6$\pm$0.7)$\times 10^{21}$ cm$^{-2}$ 
%when fitted with a power law model resulting in $\Gamma=2.6$ (Campana 2015, GCN 18483). 
The redshift $z=0.81$ was %subsequently 
measured through the MgII doublet in absorption from the 
Keck/HIRES spectrum (Perley et al. 2015). 
The isotropic equivalent energy of the burst inferred from GBM spectral data analysis in 
\cite{Toelge18492} is $E_{\gamma, \rm iso}=3.98\times 10^{52}$ erg.

In this paper we have retrieved and analysed the publicly available BAT, XRT, and GBM data  and we triggered an approved proposal to perform late time ($\sim 1$ day) observations
with the XMM-{\it Newton} (\citealt{Jansen+01}) space observatory.
In the 
following sections 
%{\bf Appendix \ref{app:datared}}
we briefly describe the procedures adopted for the data 
selection/extraction and analysis. % {of the publicly available data}. 

\subsubsection{Fermi--GBM data extraction}
We selected the GBM--CSPEC data\footnote{GBM data were downloaded from the official {\it Fermi} website \url{http://fermi.gsfc.nasa.gov/}.} (1.024 s time resolution) of the brightest detectors: NaI \#\,0, NaI \#\,3, and BGO \#\,0. Data filtering, background spectrum extraction, and timeslice selection was  performed with the software \textsc{RMFIT v.4.3.2} using standard procedures (see e.g. \citealt{Nava+11,Gruber+14}). Channels with energy $\in$[10,800] keV  and $\in$[300, 2000] keV were considered for the NaI and BGO, respectively. 

GBM spectra and background files were exported to XSPEC(v12.7.1) format in order to fit them jointly 
with {\it Swift}/BAT and XRT data. 
Details on the spectral analysis and models adopted are given below (\S 3).

\subsubsection{{\it Swift}--BAT and XRT data extraction} 
We extracted and reduced the {\it Swift}/BAT spectra and light curve\footnote{The BAT event files were downloaded from {\it Swift} data archive 
(\url{http://heasarc.gsfc.nasa.gov/cgi-bin/W3Browse/swift.pl}).} 
with the {\it Swift} software included in the HEASoft package (ver.6.17), 
using standard procedures\footnote{The latest calibration files (CALDB release 2015--11--13) were adopted.}.

We retrieved\footnote{{\it Swift} Science Data Center at the University of Leicester website: \url{http://www.swift.ac.uk/xrt_curves/} (\citealt{Evans_2009}).} the {\it Swift}/XRT count rate light curve (Fig. \ref{fig:lcprompt} -- blue line) and the intrinsic and galactic extinction corrected 0.3--10 keV flux light curve (Fig. \ref{fig:lcaft} -- black symbols). We used intrinsic \nh = 4.4$\times$10$^{21}$ cm$^{-2}$ inferred from late time XRT spectra and galactic $N_{\rm H, gal}$ = 3.7$\times$10$^{20}$ cm$^{-2}$.

% and the intrinsic and galactic extinction corrected (for \nh = 4.4$\times$10$^{21}$ cm$^{-2}$
% and $N_{\rm H, gal}$ = 3.7$\times$10$^{20}$ cm$^{-2}$, respectively) 0.3--10 keV flux light curve (shown in Fig. \ref{fig:lcaft} -- black symbols) were retrieved (\citealt{Evans_2009})\footnote{{\it Swift} Science Data Center at the University of Leicester website: \url{http://www.swift.ac.uk/xrt_curves/}}.
 
 XRT spectra\footnote{The XRT event file was retrieved from the archive of the {\it Swift}/XRT website (\url{http://www.swift.ac.uk/archive/}).} were corrected for pile-up following the procedure in \cite{Romano+06}.
 Windowed Timing mode (WT) counts below 0.5 keV were excluded owing to the  abnormal photon redistribution. %under 0.5 keV 
Count spectra were rebinned 
%with the \textit{grppha} tool 
requiring a minimum of 20--30 counts per bin.

\subsubsection{XMM-{\it Newton} observation}
XMM-{\it Newton} started observing GRB 151027A starting on 2015 October 28 at 01:19:34.00 UT (21.3 hr after the burst).
The observation lasted  for 51.5 ks without interruption.
Data reduction was performed with the XMM-{\it Newton} Science Analysis Software (SAS) version xmmsas\_20131209\_1901-13.0.0
and the latest calibration files. Data were first locally reprocessed with {\tt epproc}, {\tt emproc}, and {\tt rgsproc}.
The RGS data contained too few photons and were not considered any further.
MOS and pn data were searched for high-background intervals, and none were found. EPIC data were grade filtered using pattern 0--12 (0--4) for MOS (pn) data, and {\tt FLAG==0} and \#XMMEA\_EM(P) options.
The pn and MOS events were extracted from a circular region of 870 pixels centred on source.
Background events were extracted from similar regions close to the source and free of sources.
MOS and pn data were rebinned to have 20 counts per energy bin.
MOS data were summed and fitted within the 0.3--10 keV range, pn data within the 0.2--10 keV range.

\subsection{Optical data}

The earliest optical observations (Wren et al. 2015) %, GCN 18495) 
started with the RAPTOR 
network of robotic optical telescopes 24 s after the trigger;   a bright optical 
counterpart ($R\sim 13.7$) was found. 
Subsequent optical/near-IR observations were performed by several ground-based telescopes. 
We have collected all the magnitudes reported in the GCN in $R$ filter of wavelength $\lambda=658$ nm (see Table \ref{tabottico} in Appendix \ref{app:data} for the calibrated and galactic extinction 
corrected, E$_{\rm B-V}=0.04$, flux light curve). 
 
These are the data we use for the modelling of the GRB emission in \S 3.3.

{\it Swift}/UVOT detected GRB\,151027A in all its photometric filters (\citealt{Balzer18502}).
We retrieved UVOT public data from the UK {\it Swift} Science Data Centre 
(\url{http://www.swift.ac.uk/archive/}) and analysed them with the standard UVOT tools distributed 
within the HEASoft (v6.17).
%Through the \textit{uvotmagsource} tool we derived the magnitudes in the six filters considering 
%an extraction region centred on the source with radius 10 arcsec. 
%The background was estimated in an annulus centred on the source with inner and outer radii of 10 and 30 arcsec, respectively.
%an extraction region centred on the source with radius 10 arcsec and estimated background far 
%from the source within a circle of radius 30 arcsec. 
The results of UVOT photometry are shown in Table \ref{tabsed}.
The intrinsic optical absorption is negligible\footnote{The 95\% C.L. upper limit
for the absorption is E$_{\rm B-V}<0.3$, with $\beta=0.75$.} and has been estimated from the spectral energy distribution presented in \cite{Cano18552} and is giving in Tab. \ref{tabsed}.

%Since the burst has been detected until the \textit{uvw2} UVOT filter (rest frame central wavelength $\lambda^\prime=106$ nm), the intrinsic optical absorption has been considered negligible.  

We have performed late time ($>19$ days) Target of Opportunity (ToO) observations in the optical 
with the Italian 3.6-m Telescopio Nazionale  Galileo (TNG) and with  the 8.4-m Large Binocular 
Telescope (LBT) that we briefly summarize below. 
%{\bf Details and results of these observations are given in table ... AGGIUNGERE QUESTA TABELLA}.

\subsubsection{TNG $\&$ LBT observations}

\begin{figure}[h]
   \centering
\includegraphics[width=\columnwidth]{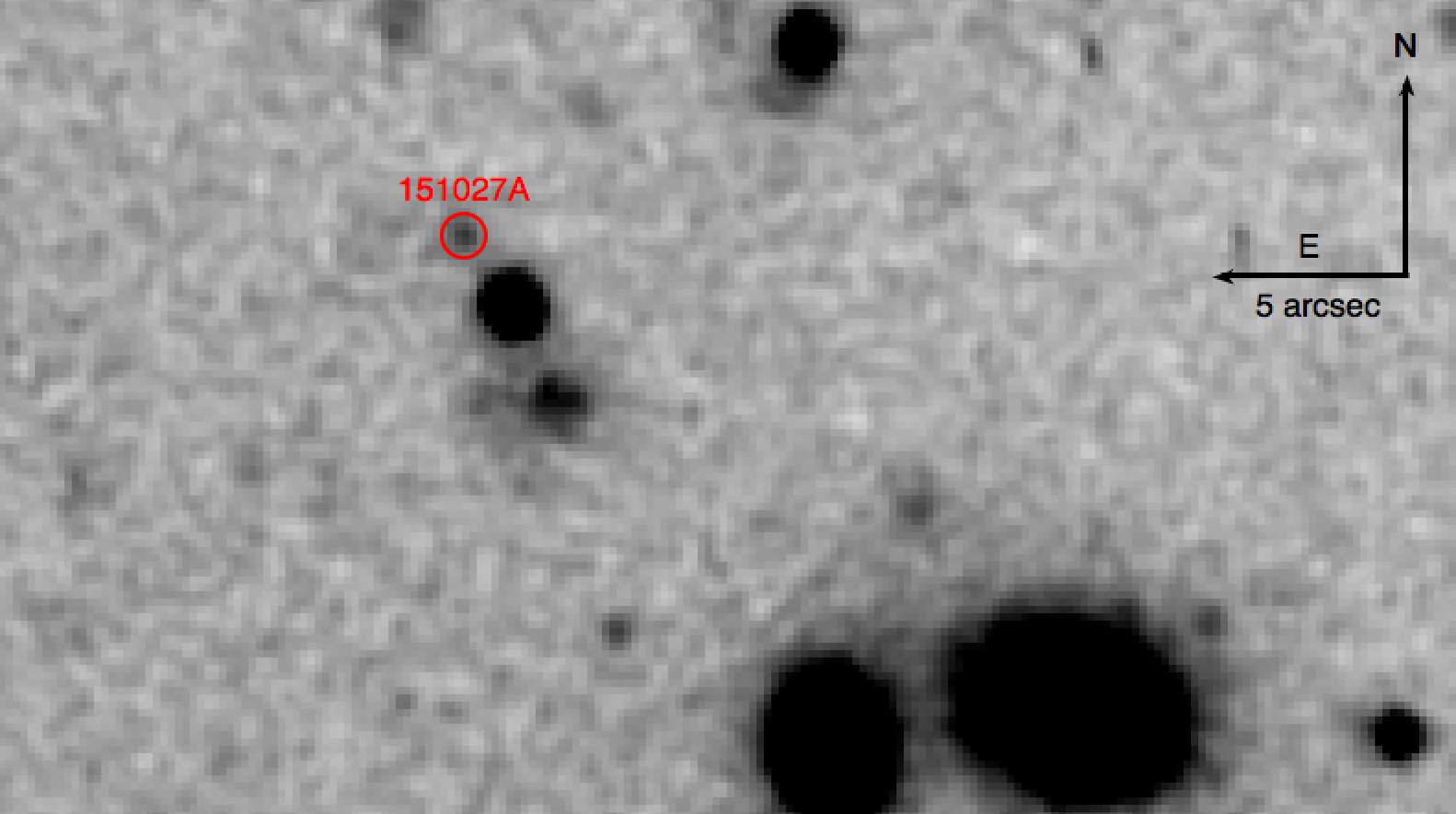}
\caption{Finding chart obtained with the TNG observation at 19 days. 
The position of the optical afterglow of GRB 151027A is shown by the red circle. 
The bright extended source in the SW (Mazaeva et al. 2015;  
Dichiara et al. 2015) is also visible. 
              }
\label{fig:fc}%
\end{figure}

We observed the optical afterglow of GRB\,151027A with TNG in the $R$ filter, for a total 
exposure of 44 min on source, starting 19.6 days after the trigger. 
Later time observations were also acquired with the 8.4 m Large Binocular Telescope (LBT) in the 
SDSS--$r$ filter, at 33.9 and 92.3 days after the event. 
The total exposure times for the LBT observations are 20 and 70 min, respectively. 
A finding chart image obtained with the TNG observation is shown in Fig. \ref{fig:fc} 
with the optical afterglow encircled.

Image reduction, including de--biasing and flat-fielding, was carried out following standard procedures. 
Images were calibrated using a set of USNO-B1 stars in the field. 
We performed point-spread function (PSF) photometry at the position of the optical afterglow 
to minimize the possible contribution of the nearby stars.

The calibrated magnitudes were corrected for the Galactic absorption along the line of sight 
(E$_{\rm B-V}$ = 0.04; \citealt{SchlaflyFink11}) and then converted into flux densities 
following \cite{Fukugita+96}. 
The results of these observations are listed at the end of Tab. \ref{tabottico}.

\subsection{Radio data}

Radio observations with the Very Large Array (VLA) 0.78 days after the trigger, performed at a mean 
frequency of 21.8 GHz, revealed  a source with flux density of $\sim$1.7 mJy (Laskar et al. 2015). 
Subsequent Giant Meter Radio Telescope (GMRT) observations (Chandra et al. 2015) 
reported a  detection most likely contaminated by a nearby bright unresolved source 
(P. Chandra - private communication). 

We triggered an approved proposal with the Medicina 32m radio telescope and obtained ToO observations with the European VLBI Network (EVN), the Very Long Baseline Array (VLBA), and the 64m Sardinia Radio Telescope (SRT, \citealt{Bolli+15}; Prandoni et al. in prep.). %, which was undergoing its astronomical validation phase. 
Details of the data acquisition and reduction are given below. 
Results of the radio observations are listed in Table \ref{tabradio}.

\subsubsection{European VLBI (EVN) observations}

We observed GRB\,151027A with the European VLBI Network on 2015 November 18 and 2016 March 15. The participating stations were Effelsberg (100 m), Medicina (32 m), Torun (32 m), Yebes (40 m), 
Westerbork (25 m), Onsala (25 m), and Jodrell Bank (25 m). 
The observing frequency was centred at 4.98 GHz, with $8\times 16$ MHz baseband channels, 
in dual polarization. 

Data were electronically transferred to the central correlator at JIVE via the so-called e--VLBI 
technique and processed in real time with the software correlator SFXC (\citealt{Szomoru2008,Keimpema2015}).
We observed in phase reference mode, alternating 1-minute scans on the nearby ($d=0.5^\circ$) 
calibrator J1806+6141 to 2.5-minute scans on the target for a total integration 
time on target of $\sim4.2$ hours. 
We also regularly observed the two check sources J1815+6127 ($d=0.7^\circ$) and J1746+6226 ($d=3.0^\circ$).
We carried out a standard calibration in AIPS, determining amplitude coefficients from gain 
curves and system temperatures recorded during the observation. 
We removed phase offsets and phase delays and rates using the phase calibrator J1806+6141. 
Phase solutions were then transferred to the target.
After applying the calibrations, we produced a dirty image of the sky which immediately 
showed a point-like source.  
We then cleaned the image. 

\begin{figure}
\centering
\includegraphics[width=\columnwidth]{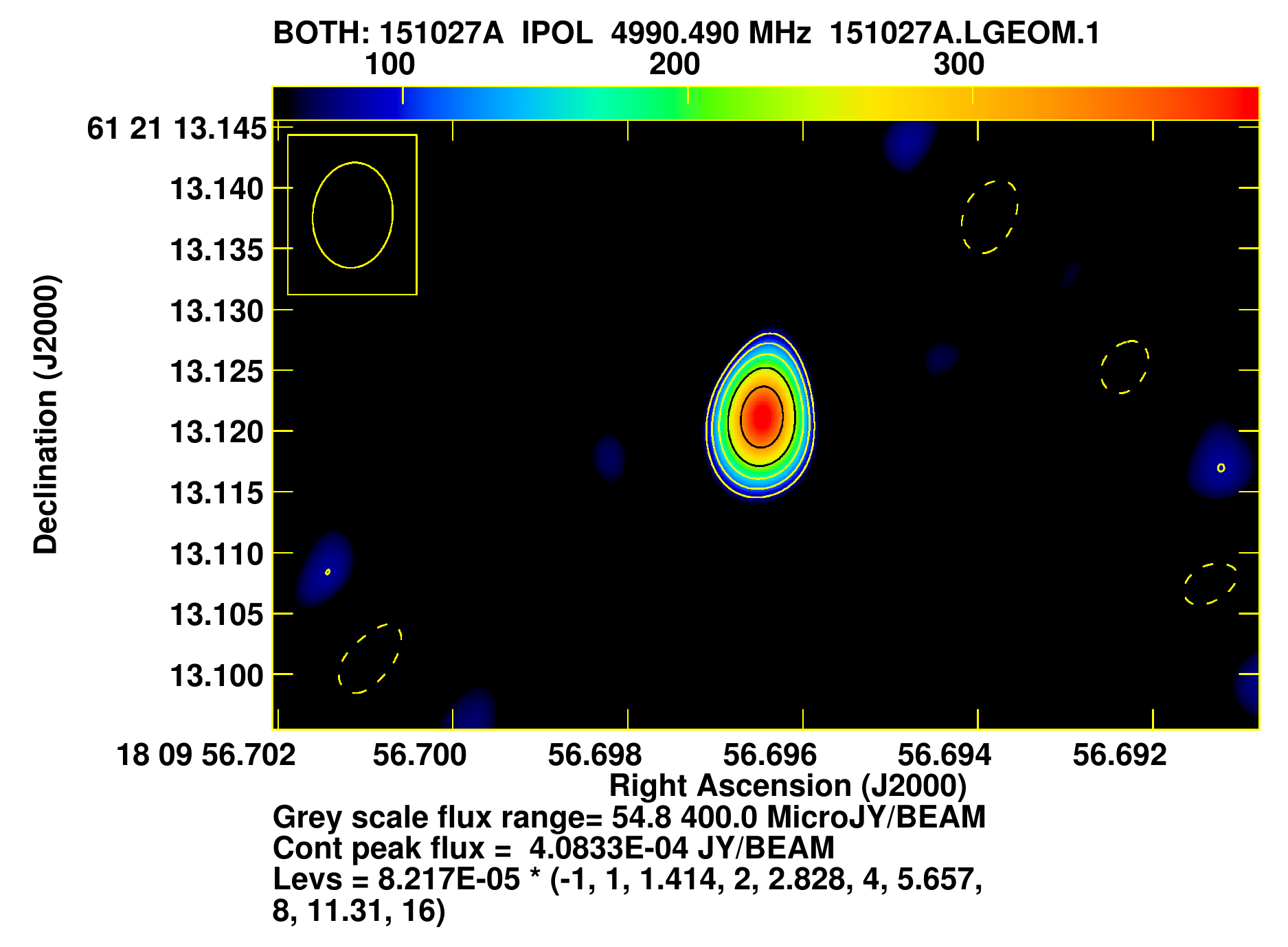}
\caption{5 GHz VLBI image of GRB 151027A taken with the EVN on 2015 November 18. 
\label{f.vlbi}}
\end{figure}

For the 2015 November observations, we achieve a noise level of 28 \mujb. A model-fit to the image plane 
with the AIPS task {\tt JMFIT} yielded the following parameters for the source: 
RA\ 18h 09m 56.6965s $\pm 0.0001$s, 
Dec.\ +61$^\circ$ $21 \arcmin \ 13.1210 \arcsec  \pm 0.0002\arcsec$, peak brightness $(400\pm50)$  \mujb. 
The component is unresolved, which implies a conservative upper limit to its size of about 1 milliarcsecond. The image is shown in Fig. \ref{f.vlbi}

For the 2016 March observations, we achieve a noise level of $22\,\mu$Jy beam$^{-1}$, and the image plane model-fit results are RA 18h 09m 56.6965s $\pm$ 0.0001s, Dec. +61$^\circ$ 21\arcmin 13.1219\arcsec $\pm$ 0.0004\arcsec, peak brightness (125 $\pm$ 15) $\mu$Jy beam$^{-1}$.

\subsubsection{Very Long Baseline Array (VLBA) observations}

We observed GRB\,151027A with the Very Long Baseline Array (VLBA) 2016 Jan 24, Jan 30, and Feb 6. 
The observations were carried out at 5, 8, and 15 GHz, with a time partition given in Table \ref{t.logvlbi}. 
At each frequency, our set-up consisted of $8\times 32$ MHz baseband channels in dual polarization. 
We used the same calibrator--target scheduling pattern as in the EVN observations, 
with a duty cycle depending on the observing frequency (see Table \ref{t.logvlbi}). 
We carried out the standard calibration in AIPS, following the latest guidelines for VLBA 
amplitude calibration. 
We combined the DBCON data with the data from observations at the same frequency taken in 
different runs; no significant variability is expected on timescales shorter than a week. 
At 5 and 8 GHz, we clearly reveal a compact source at the image phase tracking centre 
determined from the EVN observations; we also detect the source at 15 GHz, but only 
if we use natural weights in the imaging process. 
%The image is shown in Fig. \ref{f.vlbi}.

\subsubsection{Medicina observations}\label{sec:Med}
We observed GRB\,151027A with the 32m Medicina radio telescope on 2015 October 31. We observed with the Total Power backend at a central frequency of 24.5 GHz, with a bandwidth of 2 GHz. We performed 1043 cross-scans in right ascension and declination, centred on the position reported by Laskar et al. (2015). The total effective on-source time is 27 minutes. Data were calibrated with scans on NGC7027, and sky opacity was determined and compensated for through regular (about one per hour) skydip scans. No significant emission was detected above a 3 $\sigma$ noise level of 8.0 mJy.

\subsubsection{Sardinia Radio Telescope (SRT) observations}
We observed GRB\,151027A with the 64m Sardinia radio telescope on 2015 October 30-31 between 23:30 UT and 01:30 at 22 GHz, and between 01:30 and 04:30 at 7.2 GHz. The observing strategy was based on cross-scans in azimuth and elevation directions, with the following parameters for each band: at 22 GHz we observed with a bandwidth of 2 GHz with a total of 242 scans for an effective on-source time of 5 minutes; at 7.2 GHz the bandwidth was 680 MHz with 336 scans and a net on-source time of 14 minutes. Owing to scheduling constraints, the observations took place at low elevation, between about 12$^\circ$ and 19$^\circ$. No significant emission was detected above a 3 $\sigma$ noise level of 6.0 mJy and 0.6 mJy respectively at the two frequencies. These values are dominated by the low elevation at high frequency and by confusion at low frequency.
%We observed GRB\,151027A with the 64m Sardinia radio telescope on 2015 October 30. 
%At the time of observations, the Sardinia radio telescope was undergoing its astronomical validation. 
%We adopted an observing strategy similar to that of the Medicina observations. We observed at a central frequency of 22.0 GHz; the cross scans were carried out in azimuth and elevation directions; the total number of scans was 242 and the total effective on source time was 5 minutes; all other parameters were the same as in \S \ref{sec:Med}. Due to scheduling constraints, the observations took place at low elevation, starting at 17.5$^\circ$ and finishing at 12.9$^\circ$. No significant emission was detected above a 3 $\sigma$ noise level of 6.0 mJy.

\begin{table}
\centering
\begin{tabular}{cccc}
 \hline 
  Band &  
  $t-t_0$ [s]&
  $F_{\nu}$ [mJy]& Ref.
   \\
  \hline \hline
1.4 GHz & $1.64\times 10^6$ &  0.312	$\pm\,$ 0.064 & [1] \\
5 GHz & $1.92\times 10^6$ &  0.39	$\pm\,$ 0.05 & [2] \\
5 GHz & $7.69\times 10^6$ &  0.29	$\pm\,$ 0.05 & [3] \\
5 GHz & $1.20\times 10^7$ & 0.15	$\pm\,$ 0.03 & [2] \\
7 GHz & $3.31\times 10^5$ &	$< 0.6$ & [4] \\
8.4 GHz & $7.95\times 10^6$ &  0.18	$\pm\,$ 0.03 & [3] \\
15 GHz & $8.51\times 10^6$ &  0.14	$\pm\,$ 0.03 & [3] \\
22 GHz & $6.74\times 10^4$ & 1.7 & [5] \\
22 GHz & $3.31\times 10^5$ & $<6.0$ & [4] \\
24 GHz & $3.92\times 10^5$ & $<8.0$ & [6] \\

\hline \hline
\end{tabular}
\caption[]{Radio band fluxes and 3$\sigma$ upper limits used in this work.}

\tablebib{[1]~\citet{Chandra18608}; [2] this work: EVN observations; 
[3] this work: VLBA observations; [4] this work: SRT observations;
[5] \citet{Laskar18508}; [6] this work: Medicina observations;
}
\label{tabradio}
\end{table}

%______________________________________________

\section{Data analysis and results}

\subsection{Prompt emission: First and second peaks}\label{sec:he_analysis}

During the first two peaks of the light curve, corresponding to the time interval 
$\sim$0-24 seconds, we extracted three spectra: \#1 and \#2 corresponding respectively to the rise and 
decay phase of the first peak and \#3 for the entire duration of the second (dimmer) peak (referring to the labelled regions in the middle panel of Fig. \ref{fig:lcprompt}). 
We jointly fit the {\it Fermi}/GBM (NaI and BGO) and the {\it Swift}/BAT spectra 
with a cutoff power law model (CPL) with a free normalization constant between {\it Fermi} 
and {\it Swift}. 
Start and stop times and the best fit parameters (with 68\% confidence errors) and the 
$\chi^2$(dof) are given in Table \ref{tab:prompt}. 
Spectrum \#3 can be fitted only with a simple power law model (i.e. the $E_{\rm peak}$ 
of the cutoff power law model is unconstrained). 
These three spectra are shown in the top panels of Fig. \ref{fig:lcprompt} where the 
data (green and red for the GBM and BAT, respectively) and the best fit model 
(solid black line) are shown.

\begin{table*}
\center
\footnotesize
\begin{tabular}{llllllllllll}
\hline
Data$^{a}$ & \#& start$^{b}$ & stop$^{b}$  & Model$^{c}$ & $\Gamma$ & $E_{\rm peak}$ & A$^{d}$ & kT  & A$_{\rm BB}$ & $\chi^{2}$(dof) & P$^{\rm F-test}_{CPL-CPLBB}$\\

 & & s & s &  & & keV & ph cm$^{-2}$ s$^{-1}$ & keV & ph cm$^{-2}$ s$^{-1}$ & &\\
\hline 
\hline
B+G & 1& -0.256	& 1.792	& CPL & 0.92$^{+0.08}_{-0.08}$ & 207$^{+29}_{-23}$ & 2.93$^{+0.84}_{-0.68}$  & - & - & 164(244) & -  \\
... & 2 & 1.792	& 6.912	& CPL & 1.29$^{+0.15}_{-0.13}$ & 69$^{+16}_{-6}$ & 6.09$^{+2.84}_{-1.38}$  & - & - & 192(272) & - \\
...	& 3 & 17.152	& 23.296	& PL & 1.82$^{+0.08}_{-0.08}$ & - &  9.43$^{+3.26}_{-2.43}$  & - & - & 168(286) & - \\

\hline
X+B+G & 4 & 90 & 100 & CPL+BB & 1.26$_{-0.05}^{+0.04}$ & 218$_{-59}^{+103}$ & 0.71$_{-0.10}^{+0.09}$  & 1.23$_{-0.19}^{+0.33}$ & 0.041$_{-0.008}^{+0.009}$ & 211(256) &  3.8$\times10^{-7}$ \\
... & 5 & 100 & 110 & CPL+BB & 1.06$_{-0.02}^{+0.02}$ & 316$_{-41}^{+51}$ & 0.89$_{-0.06}^{+0.06}$  & 3.02$_{-0.20}^{+0.21}$ & 0.17$_{-0.02}^{+0.02}$ & 279(301) &  8.0$\times10^{-13}$ \\
... & 6 & 110 & 120 & CPL+BB & 1.18$_{-0.03}^{+0.02}$ & 209$_{-22}^{+27}$ & 2.08$_{-0.16}^{+0.16}$  & 2.01$_{-0.08}^{+0.07}$ & 0.55$_{-0.04}^{+0.05}$ & 257(296) &  6.3$\times10^{-35}$ \\
... & 7 & 120 & 130 & CPL+BB & 1.50$_{-0.06}^{+0.05}$ & 76$_{-14}^{+22}$ & 2.71$_{-0.37}^{+0.37}$  & 1.12$_{-0.06}^{+0.07}$ &  0.51$_{-0.03}^{+0.03}$ & 284(293) & 8.7$\times10^{-45}$ \\

\hline
X & 8 & 130 & 140 & BB & - & - & -  & $0.63^{+0.02}_{-0.02}$ & $0.25^{+0.01}_{-0.01}$ & 39(34) & - \\
... & 9 & 140 & 150 & PL+BB & $1.53^{+0.27}_{-0.38}$ & - & $1.19^{+0.40}_{-0.47}$& $0.43^{+0.04}_{-0.03}$ & $0.087^{+0.015}_{-0.013}$ & 30(26) & 2.3$\times10^{-6}$ \\
... & 10& 150 & 160 & PL+BB & $2.07_{-0.23}^{+0.19}$ & - & $0.86_{-0.21}^{+0.18}$  & $0.37_{-0.03}^{+0.03}$& $0.037_{-0.005}^{+0.006}$ & 36(46) & 4.6$\times10^{-10}$ \\
... & 11 & 160 & 170 & PL & $2.53_{-0.06}^{+0.06}$ & - & $1.19_{-0.04}^{+0.04}$  & - & - & 53(45) & (0.02)$^e$ \\
... & 12 & 170 & 180 & PL & $2.45_{-0.06}^{+0.07}$ & - & $1.01_{-0.04}^{+0.04}$  & - & - & 50(40) & (0.01)$^e$ \\
... & 13 & 180 & 190 & PL & $2.65_{-0.08}^{+0.09}$ & - & $0.76_{-0.03}^{+0.09}$  & - & - & 16(30) & - \\
... & 14 & 190 & 200 & PL & $2.51_{-0.10}^{+0.10}$ & - & $0.57_{-0.03}^{+0.03}$  & - & - & 25(22) & - \\

\hline
XMM & & 7.8$\times 10^5$ & 1.3$\times 10^6$ & PL+BB & $2.09^{+0.03}_{-0.04}$ & - & $3.8^{+0.1}_{-0.2}\times 10^{-4}$& $0.11^{+0.03}_{-0.02}$ & $3.1^{+2.0}_{-1.1}\times 10^{-6}$ & 398(345) & 5.8$\times10^{-7}$ \\
%XMM & & 7.8$\times 10^5$ & 1.3$\times 10^6$ & PL+BB & $2.09^{+0.03}_{-0.04}$ & - & $3.79^{+0.11}_{-0.15}\times 10^{-4}$& $0.11^{+0.03}_{-0.02}$ & $3.05^{+1.99}_{-1.13}\times 10^{-6}$ & 398(345) & 5.8$\times10^{-7}$ \\
\hline\hline
\end{tabular}
\caption{Prompt emission time resolved spectral analysis. 
$^{a}$Spectral data used in the fit: B=Swift/BAT, G=Fermi/GBM and X=Swift/XRT. 
$^{b}$Times refer to the trigger time of the burst. 
$^{c}$Models adopted in the fit: CPL=powerlaw with exponential cutoff, PL=powerlaw, BB=blackbody; galactic ($N_{\rm H, gal}$=3.7$\times$10$^{20}$ cm$^{-2}$) and intrinsic (\nh =4.4$\times$10$^{21}$ cm$^{-2}$) absorption is present in all models (using Tuebingen--Boulder ISM absorption model, \citealt{Wilms+00}).
$^{d}$Spectral normalization is computed at 1 keV. 
$^e$In this case the addition of a BB component is not statistically significant, as suggested by the value of the null hypothesis probability associated with the F--test.
The horizontal lines correspond to the differently shaded regions in Fig. \ref{fig:lcprompt}
% first two peaks (which used only GBM and BAT spectra) from the 
%third one which includes the XRT data in the spectral analysis. 
%{\bf AGGIORNARE CON I NUOVI VALORI}
\label{tab:prompt} }
\end{table*}

\subsection{Evidence of a thermal component: Third peak}

The third peak of the light curve was observed by BAT and GBM above 10 keV and 
simultaneously by XRT in the 0.5 -- 10 keV energy range. 
The light curves (see middle panel of Fig. \ref{fig:lcprompt}) show that the 
XRT peak is delayed with respect to that observed by BAT and GBM. 
We selected four time intervals (from 90 s to 130 s after the trigger) where the data from three instruments  overlap, and jointly fitted the spectra.
This allows us to perform a time resolved spectral analysis over a wide energy range, namely from 0.5 keV to a few MeV. 

We fit the spectra with a CPL model.
Since the data extend down to 0.5 keV, it is necessary to take into account  the galactic and intrinsic
absorption. The Tuebingen--Boulder ISM absorption model (\citealt{Wilms+00}) encoded in the {\it tbabs} and {\it ztbabs} models of XSPEC is used. We assume the galactic absorption $N_{\rm H, gal}=3.7\times 10^{20}$ cm$^{-2}$ and keep it fixed, and   we also allow for an intrinsic (at $z$=0.81) absorption. Also in this case we allow a free normalization constant between the {\it Swift}/BAT spectrum 
and the {\it Fermi}/GBM (NaI+BGO) spectra. In all the fits we find that this constant is within  a factor of 2 and is consistent with 1.0.

 If the intrinsic \nh\ is treated as a free parameter,
we find that it varies dramatically (by more than one order of magnitude) describing a peak over a 30-second timescale coincident with the flare. We interpret this non-physical feature as being indicative of the possible presence of an additional component during the flare.
We therefore fixed the intrinsic \nh=0.44$\times 10^{22}$ cm$^{-2}$ 
which is the value found by fitting the XRT spectra at very late times (i.e. $>$5 days). 

By visual inspection of the fitted spectra and their residuals we noticed systematic deviation from the model in the XRT 0.5-10 keV energy range, making the CPL fit unacceptable.
We therefore tried to model this excess by adding the simplest two-parameter thermal blackbody  (BB) component. 
We refitted the data and compared the new fit (i.e.  absorbed cutoff power law plus 
blackbody -- CPL+BB) with the old one (i.e. absorbed CPL) 
through an F--test. 
We find that in all of the four spectra describing the third emission episode of GRB\,151027A there 
is  statistically significant evidence for the presence of a thermal blackbody component. 
The probability of the F--test (representing the probability that the fit is not significantly improved by the additional BB component) 
is given in Table
\ref{tab:prompt}, along with the spectral parameters of the CPL+BB fit. 
The four spectra are shown in the bottom panels of Fig. \ref{fig:lcprompt}.

The addition of the BB component to the CPL is the minimal assumption that can produce a curvature 
of the spectrum which adapts to the data points. 
However, we also verified whether the systematic deviation of the data from a simple CPL could also be 
accommodated  by a second CPL. 
In order to have a similar number of free parameters of the BB, in this case we fixed the second 
CPL low energy photon index to the value predicted for single electron synchrotron emission, i.e. 2/3. 
In  spectra \#3 and \#4, when the peaked component
is less dominant, the fits performed using CPL+CPL or CPL+BB 
are statistically equivalent. 
Afterwards, when the component at low energies represents a considerable fraction of the total flux, 
the CPL+BB model is statistically preferable.

\subsection{X--ray emission in the interval 130--200 seconds}
After 130 s, the GBM and BAT data cannot be used for the spectral analysis.
We analyse seven XRT spectra (corresponding to 
%the last 7 hatched region 
regions 8--14
in the middle panel of Fig. \ref{fig:lcprompt}) in the time interval 130--200 s and fit with %a we can use only the XRT data for the  which can be fitted with 
an absorbed 
power law (PL) or an absorbed power law plus a blackbody component (PL+BB).
Given the limited energy range 0.5--10 keV we cannot determine the peak of a possible cutoff power law model.
For each spectrum the statistical significance of the addition of the thermal component has been estimated through the F--test.
 For  spectrum \#8 %(between 130 and 140 s) 
%is not possible to constrain the parameters of the power law component and 
the best fit is obtained with a pure BB model since the addition of a power law component does not  constrain the power law fit parameters. %the quality of the fit, ${\rm P}^{\rm F-test}_{\rm BB-PLBB}=$????).
In the following spectra the best fit model is PL+BB, in which the thermal component remains statistically significant up to 160 s. After that, the spectrum is best fitted by a single PL component.

\begin{figure}
\centering
%\hspace{-3em}
%\includegraphics[width=\columnwidth]{timevo_plot_bis.eps}
\includegraphics[width=\columnwidth]{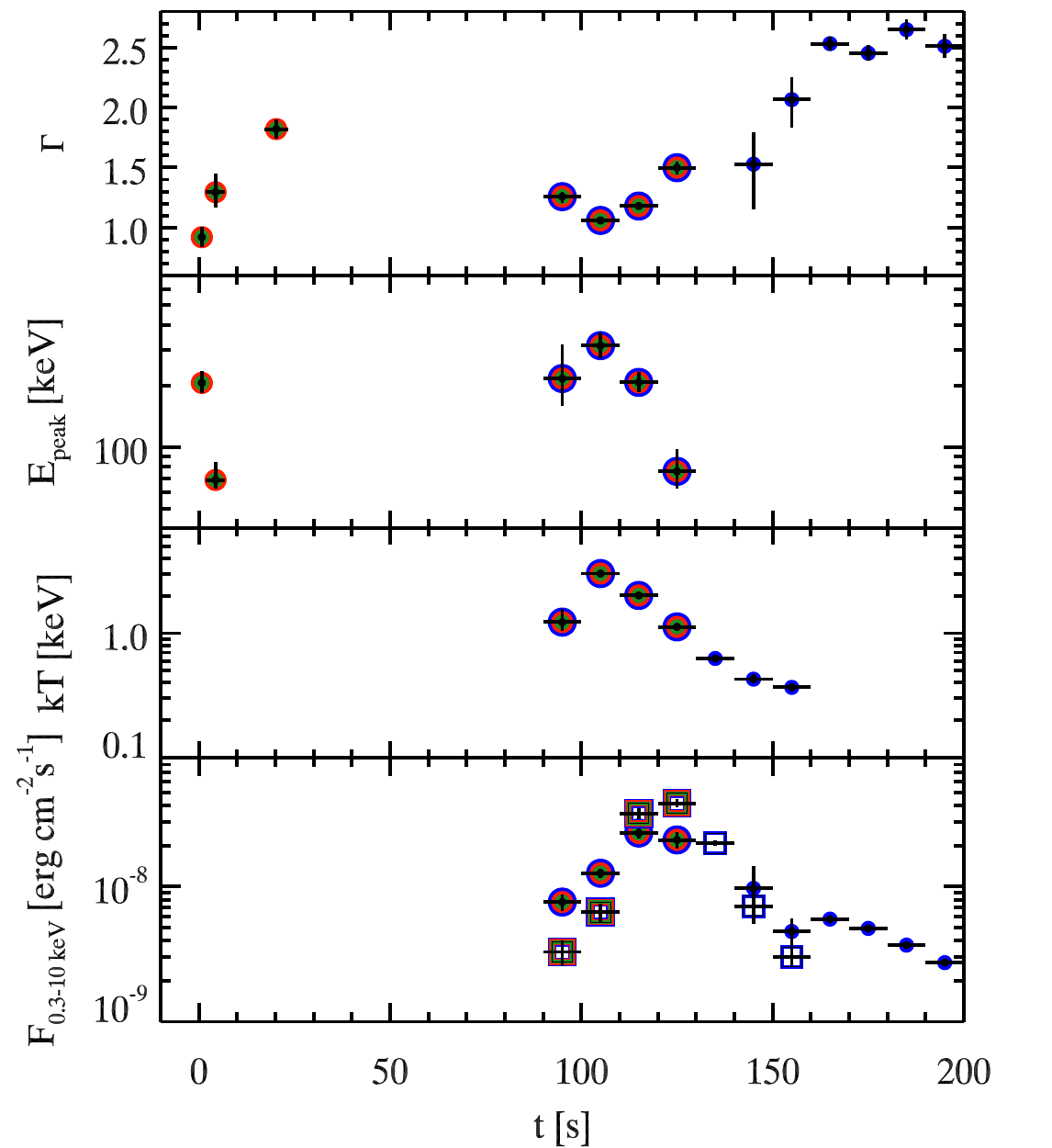} 
\caption{Spectral evolution of GRB 151027A parameters through the entire burst duration. 
Panel (a) shows the temporal evolution of the non-thermal component spectral index.
Panel (b) shows the evolution of the peak energy when the non-thermal component is described by a CPL. 
Panel (c) shows the BB temperature evolution. 
The last panel (d) shows the comparison between the fluxes, integrated in the 0.3--10 keV energy range, 
associated with the thermal (squares) and non-thermal component (points). The colour codes of the different symbols corresponds to the spectral data sets (the same colour coding) shown in Fig.\ref{fig:lcprompt}. 
}
\label{fig:parflare}%
\end{figure}

The evolution of the spectral parameters is shown in Table \ref{tab:prompt} and Fig. \ref{fig:parflare}. 

\begin{figure}
\centering
   \includegraphics[width=\columnwidth]{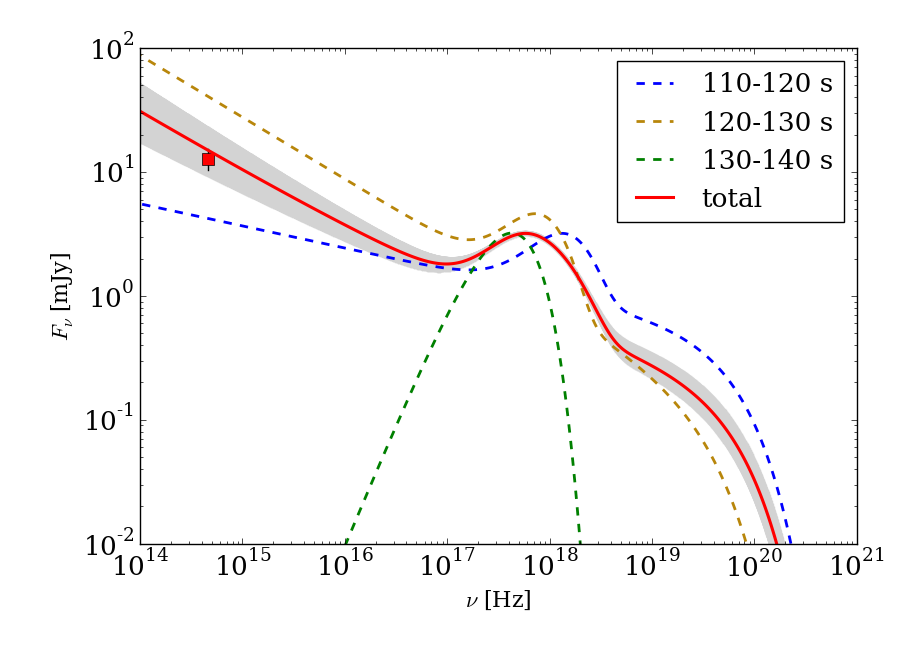}
\caption{Comparison between the optical $R$ band detection at 126 s (red square, Pozanenko et al. 2015) and the 
fit of the composite XRT-BAT-GBM spectrum (red solid line). 
The grey shaded area represents the 1$\sigma$ scatter from the best fit.}
\label{fig:speflareopt} 
\end{figure}

The results of the BAT-GBM-XRT spectral fits were compared with the optical $R$ band detection 
at 126 s (\citealt{Pozanenko18635}). 
The optical detection is compatible with the low energy extrapolation of the model (Fig. \ref{fig:speflareopt}). 
This result suggests that the early optical emission could be produced by the same mechanism responsible 
for the high energy emission and therefore it should not be interpreted as standard afterglow.

\subsection{XMM-Newton late time spectrum}
The XMM-{\it Newton} late time spectral analysis was intended to obtain a more accurate estimate of intrinsic \nh. We initially performed the fit using a PL model with free intrinsic absorption. From the residual, we noticed that a peaked component should be added to improve the fit. For this reason we refit the spectrum using a PL+BB model with free absorption. The XMM-{\it Newton} spectrum showed a still statistical significant thermal component that contributes to only $\sim$8\% of the 0.3--10 keV flux. The BB temperature was lower than the one obtained from  XRT spectrum \# 10 (the last time interval where BB was detected).
All the fit parameters are listed in Table \ref{HEdata}.

The best fit \nh \ parameter obtained in the PL+BB model is fully compatible with the value obtained by the late time XRT spectrum\footnote{In particular, from the XMM-{\it Newton} spectrum we obtained \nh=$(0.42\pm 0.05)\times 10^{22}$ cm$^{-2}$}. 

The 0.3--10 keV flux corrected by intrinsic and galactic absorption is compatible with the flux measured by XRT at that time and it is shown by the light blue diamond in Fig. \ref{fig:lcaft}.

\subsection{Radio}
As an example of the VLBI data quality, we show in Fig. \ref{f.vlbi} the EVN image at 5 GHz
%VLBA image at 8 GHz
obtained %with the combination of the datasets taken 
on 
2015 Nov 18.
%2016 Jan 24 and 30. 
We give in Table \ref{t.resvlbi} (see Appendix \ref{app:data}) the basic parameters (noise $\sigma_\nu$, peak surface 
brightness $B_\nu$, peak-to-noise ratio, and image resolution) of this and the other images; 
in Col. 7 we  list the total flux density $S_\nu$ obtained from a visibility 
model--fitting carried out in Difmap. 
Estimating the accuracy of the amplitude scale for VLBI data is traditionally a difficult task. 
From an inspection of the data quality and of the calibrator images, and taking into account the 
local noise, we conservatively estimate it to be within 20\%.

From the comparison of the EVN and the VLBA 5 GHz data, we find that the source flux density 
 decreased by nearly 50\% from day 22 to day 89, and by a further $\sim50\%$ between day 89 and 140. Moreover, from the comparison of the nearly simultaneous VLBA multi--$\lambda$ data, we determine that 
the emission region is optically thin, with a spectral index of about $\beta=0.7-0.9$, 
assuming $F_\nu \sim \nu^{-\beta}$ (see fourth panel of Fig. \ref{fig:SedAft}).

The position of the source is consistent among the various experiments to  within about 1 milliarcsecond. 
The mean coordinates are r.a.  18h 09m 56.6964s, dec.\ $+61^\circ\ 21\arcmin\ 13.1210\arcsec$. 
A more accurate astrometric calibration is beyond the scope of the present paper. 

% coordinate
% EVN 5 GHz 18 09 56.6965 61 21 13.1210
% VLBA 5GHz 18 09 56.6963 61 21 13.1216
% VLBA 8GHz 18 09 56.6964 61 21 13.1207
%A model-fit to the image plane with the AIPS task {\tt JMFIT} yielded the following parameters for the source: r.a.\ 18h 09m 56.6965s $\pm 0.0001$s, dec.\ $+61^\circ\ 21\arcmin\ 13.121007\arcsec\ \pm 0.0002\arcsec$, flux density $(400\pm50)$\mujb. The component is unresolved, which implies a conservative upper limit to its size of about 1 milliarcsecond.

\subsection{Afterglow light curve and spectral energy distributions}\label{sec:Sed}

\begin{figure*}
\centering
\vskip 0.85truecm
\includegraphics[trim={0 0 0 1.5cm},width=165mm]{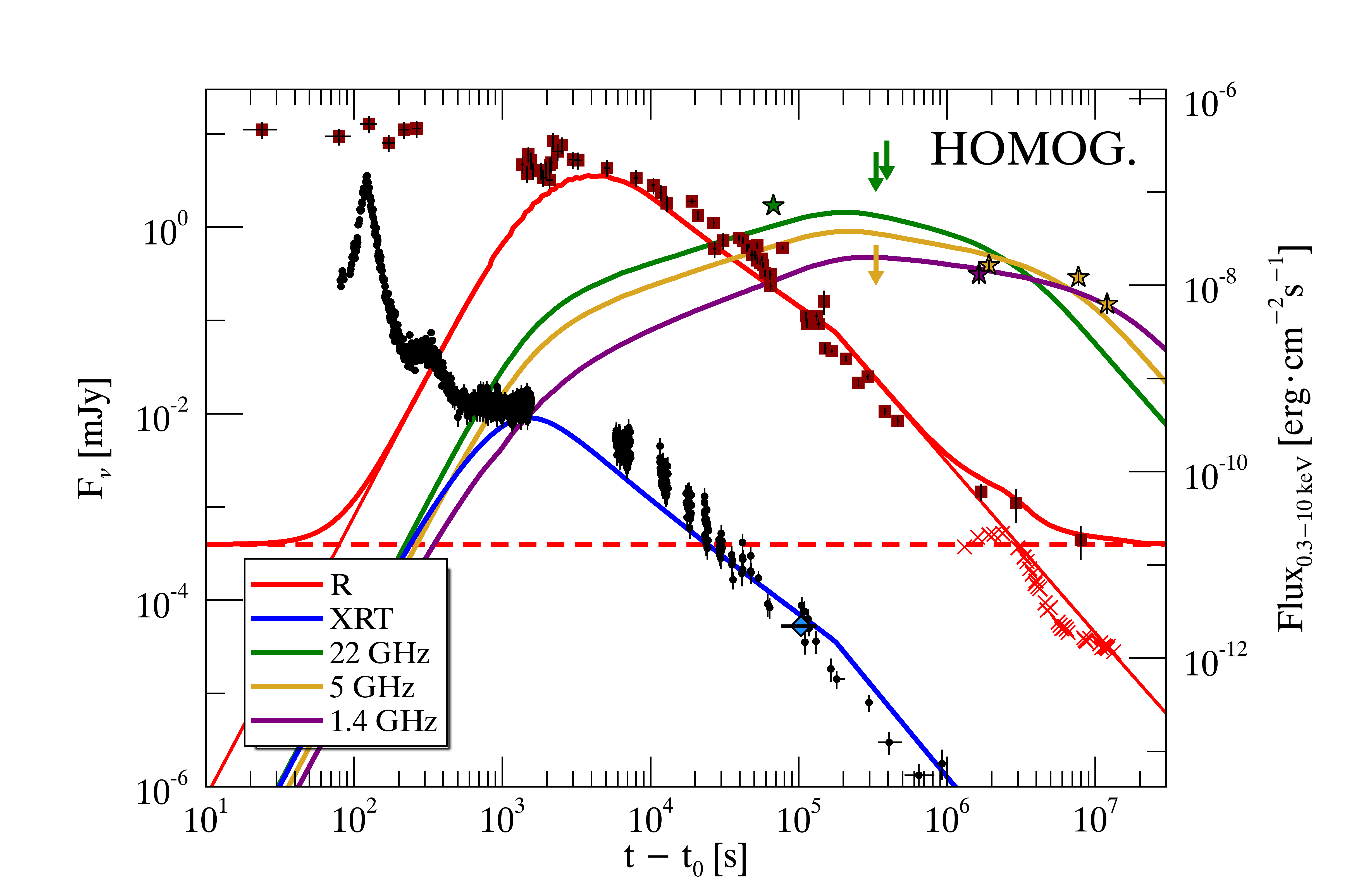}
\vskip -0.15truecm
\includegraphics[width=165mm]{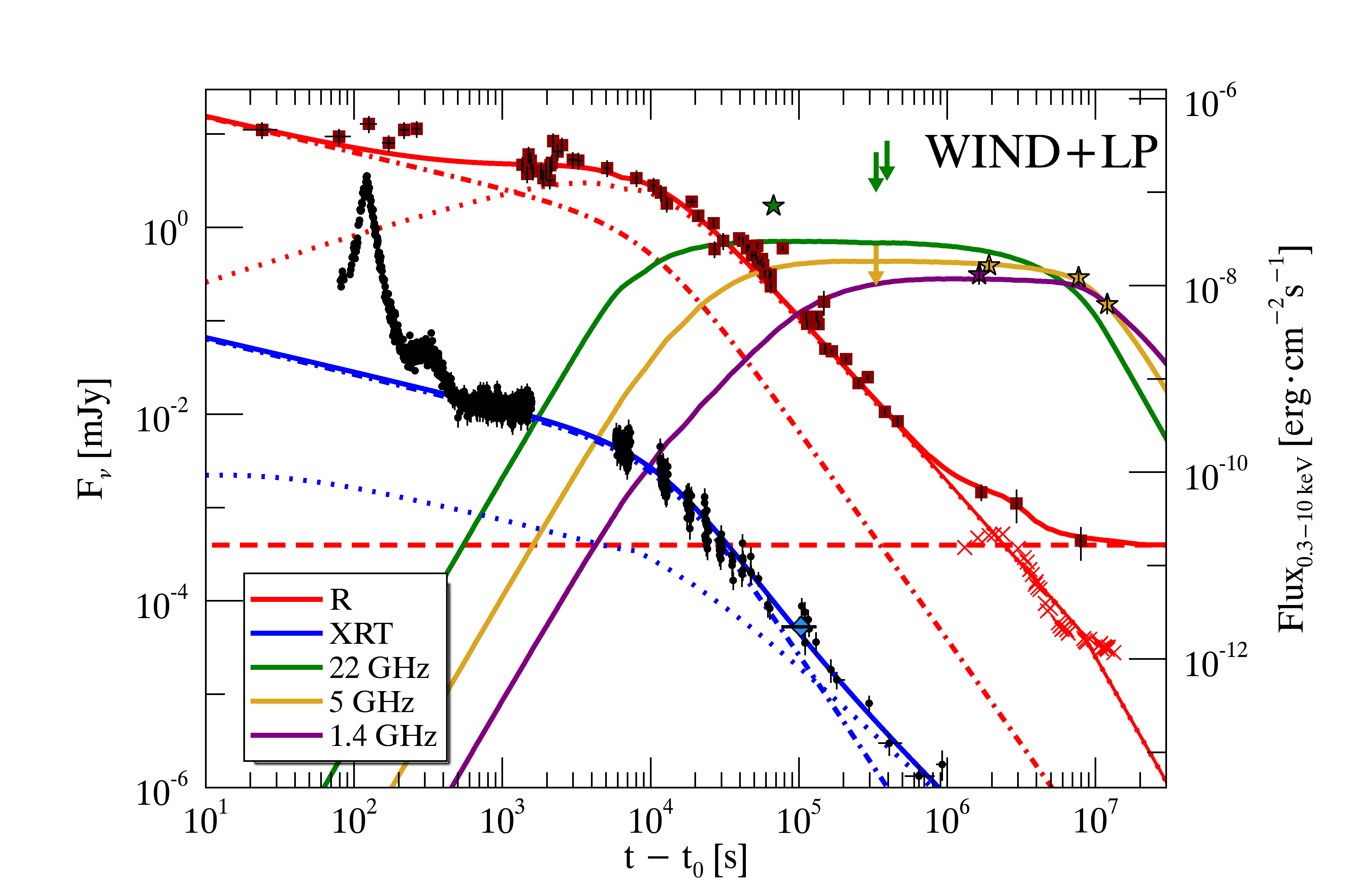}
\caption{Multiwavelength light curve of the afterglow of GRB 151027A: XRT flux in band 
0.3--10 keV (black circles, blue solid lines  referring to the right y--axis), 
$R$ band flux density $F_{\nu}$ (red squares, red solid lines), 22 GHz (green star, green solid lines), 
5 GHz radio detections and upper limits (orange stars, orange solid lines) and 1.4 GHz (purple star, purple solid lines).
The light blue diamond shows the 0.3--10 keV flux measured by XMM-{\it Newton}.
The red crosses represent a supernova light curve template obtained by shifting the light curve 
of SN1998bw at $z$=0.81 (\citealt{Galama+98}; \citealt{Clocchiatti+11}).
The red dashed line shows the estimated $R$-band flux density of the host galaxy inferred from 
the LBT observation 92 days after the trigger.
The 7 GHz SRT upper limit is shown in orange and compared with the 5 GHz model.
The 24 GHz Medicina upper limit is shown in green and compared with the 22 GHz model.
 Top: Best afterglow modelling with a homogeneous external medium.
Bottom: Best solution in the wind medium scenario in addition to a late prompt 
component. Dotted red and blue lines represent the afterglow forward shock emission in the $R$ and XRT band, respectively. Dash-dotted lines represent the late prompt component.
}
\label{fig:lcaft}%
\end{figure*}

The XRT 0.3--10 keV unabsorbed flux, the $R$ band observations (see Tab. \ref{tabottico}) and the 
radio detections and upper limits (see Tab. \ref{tabradio}) were used to build the 
multiwavelength light curve of the afterglow of GRB 151027A shown in Fig. \ref{fig:lcaft}.

\begin{figure*}
\centering
\includegraphics[width=170mm]{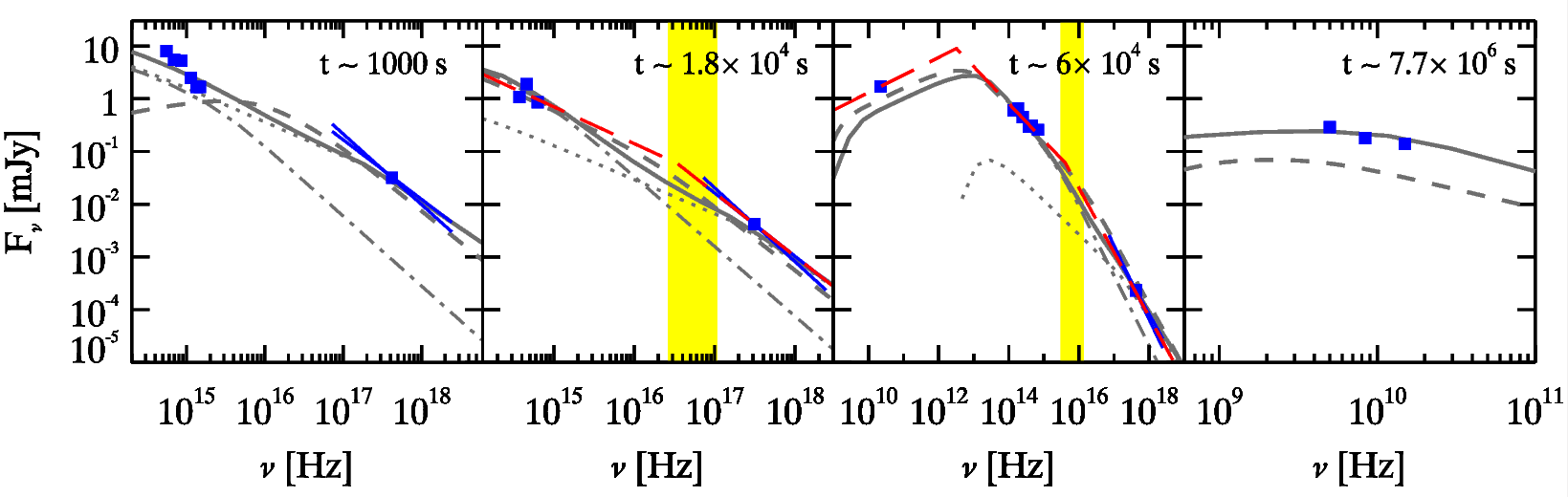}
\caption{Spectral energy distributions of the afterglow of GRB 151027A.
The four epoch data are collected in Tab. \ref{tabsed}.
The grey dashed lines represent the prediction of the model with external homogeneous medium density profile.
The grey solid lines represent the prediction of the model with wind medium and late prompt.
The afterglow component SEDs are shown by the dash-dotted lines (for the wind medium profile), while the late prompt component is shown by the dotted lines.
 The red long-dashed lines in the second and third panels show the best synchrotron spectrum compatible with the SEDs. The yellow regions show the possible positions of the cooling frequency $\nu_{\rm c}$.
}
\label{fig:SedAft}%
\end{figure*}

We built four spectral energy distributions at different times (1000 s, $1.8\times 10^4$ s, $6\times 10^4$ s, 
$7.7 \times 10^6$ s) combining the data collected from GCNs, UVOT, and XRT observations and also radio VLBA observations. 
The unabsorbed fluxes are included in Tab. \ref{tabsed} and the four SEDs are shown in Fig. \ref{fig:SedAft}.

\begin{table}
\centering
\begin{tabular}{ccccc}
 \hline 
$\nu$ [Hz] &  $F_{\nu}$ [mJy]& $\beta$ &$t_{\rm c}$ [s] & Ref.
\\
\hline \hline
$5.55 \times 10^{14}$ &  7.98	$\pm\,$0.66 & --& 1109 & [1] \\
$6.93 \times 10^{14}$ &  5.53	$\pm\,$0.25 & --& 1034 & [1] \\
$8.57 \times 10^{14}$ &  5.28	$\pm\,$0.19 & --& 893.4 & [1] \\
$1.14 \times 10^{15}$ &  2.48	$\pm\,$0.18 & --& 1072 & [1] \\
$1.34 \times 10^{15}$ &  1.66	$\pm\,$0.14 & --& 1130 & [1] \\
$1.48 \times 10^{15}$ &  1.66	$\pm\,$0.14 & --& 1086 & [1] \\
XRT & $4.9 \times 10^{-10}$ & $1.23^{+ 0.11}_{- 0.10}$ & 993 & [2] \\
\hline
$3.72 \times 10^{14}$ &  1.08	$\pm\,$0.039 & --& $1.89 \times 10^4$ & [3] \\
$4.56 \times 10^{14}$ &  1.89	$\pm\,$0.087 & --& $1.89 \times 10^4$ & [3] \\
$6.32 \times 10^{14}$ &  0.851 $\pm\,$0.047 & --& $1.89 \times 10^4$ & [3] \\
XRT & $4.3 \times 10^{-11}$ & $1.31^{+ 0.10}_{- 0.12}$ & $1.84 \times 10^4$ & [2] \\
\hline
$2.19 \times 10^{10}$ &  1.7 & --& $6.74 \times 10^4$ & [4] \\
$1.37 \times 10^{14}$ &  0.600 $\pm\,$0.055 & --& $6.34 \times 10^4$ & [5] \\
$1.84 \times 10^{14}$ &  0.658 $\pm\,$0.061 & --& $6.26 \times 10^4$ & [5] \\
$2.46 \times 10^{14}$ &  0.450 $\pm\,$0.042 & --& $6.09 \times 10^4$ & [5] \\
$3.72 \times 10^{14}$ &  0.297 $\pm\,$0.027 & --& $6.58 \times 10^4$ & [5] \\
$4.56 \times 10^{14}$ &  0.311 $\pm\,$0.034 & --& $6.35 \times 10^4$ & [6] \\
$6.74 \times 10^{14}$ &  0.258 $\pm\,$0.024 & --& $6.46 \times 10^4$ & [5] \\
XRT & $3.5 \times 10^{-12}$ & $1.24^{+ 0.18}_{- 0.10}$ & $6.35 \times 10^4$ & [2] \\
\hline
$5.0 \times 10^9$ & 0.29	$\pm\,$ 0.05 & -- & $7.69\times 10^6$ &  [7] \\
$8.4 \times 10^9$ & 0.18	$\pm\,$ 0.03 & -- & $7.69\times 10^6$ &  [7] \\
$1.5 \times 10^{10}$ & 0.14 $\pm\,$ 0.03 & -- & $7.69\times 10^6$ &  [7] \\
\hline \hline
\end{tabular}
\caption[]{Afterglow spectral energy distributions of GRB 151027A.
For the XRT data the 0.3--10 keV  flux  
in erg/cm$^2$/s is shown in addition to the spectral slope $\beta$.}

\tablebib{[1] this work: UVOT observations; 
[2] XRT automatic analysis (\url{http://www.swift.ac.uk/xrt_products/00661775}); 
[3] \citet{Yano18491}; [4] \citet{Laskar18508}; [5] \citet{Cano18552};
[6] \citet{Oksanen18567}; [7] this work: VLBA observations.
}
\label{tabsed}
\end{table}

\section{Discussion}

\subsection{Prompt emission and flare}

The prompt light curve of GRB\,151027A shows three isolated emission peaks.
The first two peaks have a standard behaviour with non-thermal spectra both characterized by a hard to soft evolution.

\begin{figure}
\centering
\includegraphics[width=\columnwidth]{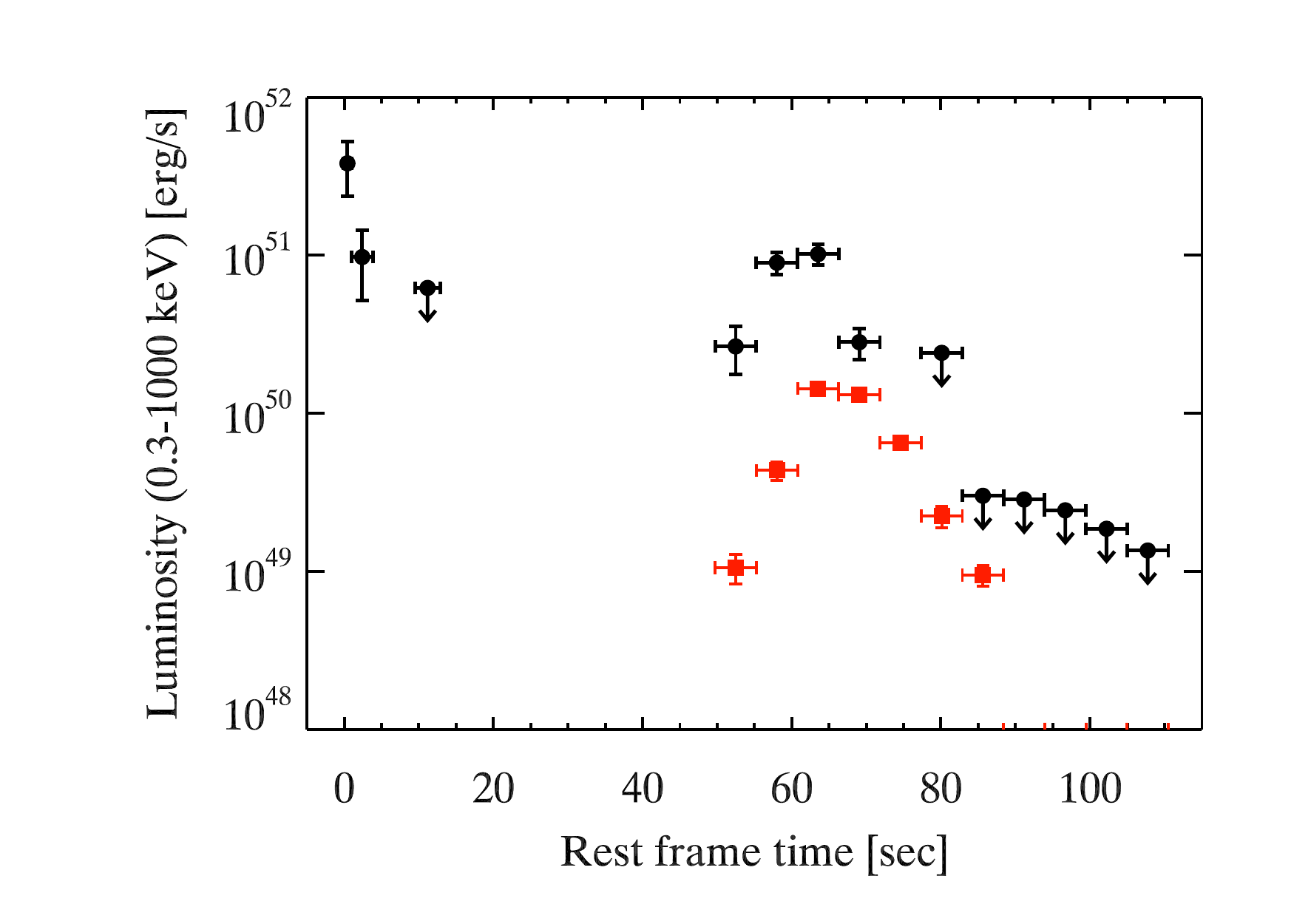}
\vskip -0.6truecm
\includegraphics[width=\columnwidth]{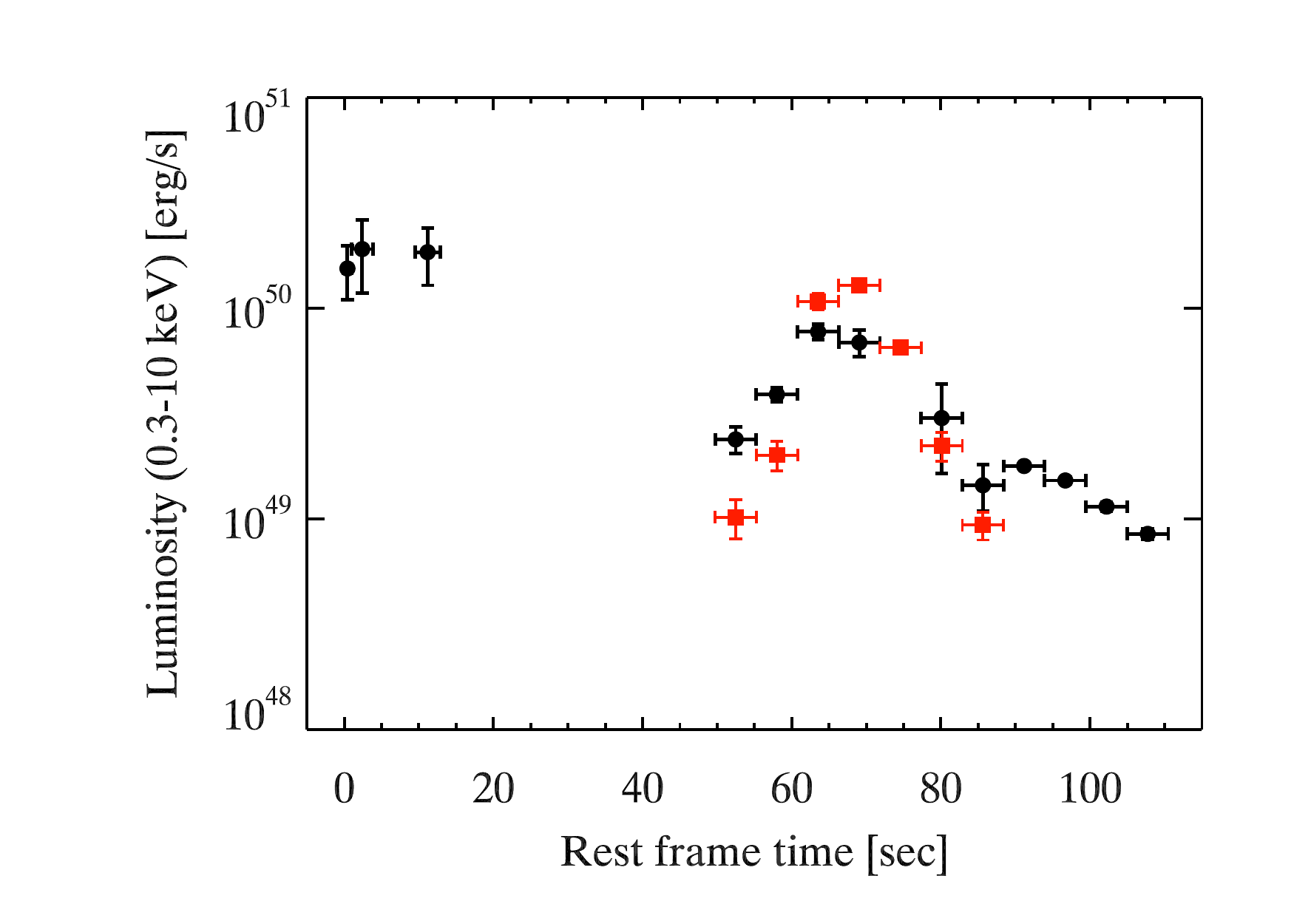}
\caption{Top: 0.3--1000 keV luminosity as a function of rest frame time divided into thermal 
(red squares) and non-thermal component (black circles). 
Upper limits to the non-thermal luminosity are shown when the peak energy of the cutoff powerlaw component is not constrained.
Bottom:  Same as the top, but for luminosity in the range 0.3--10 keV.}
\label{fig:lum_flare}%
\end{figure}

The third peak
shows a statistically significant 
 BB component at low energies 
superimposed on a cutoff power law.
Evidence of a thermal emission have also been found in other GRB spectra. 
Typically it has been detected in the early phases of the prompt emission (\citealt{Ghirlanda+03}) 
or it can be present throughout the entire burst duration (\citealt{Ryde04,Bosnjak+06,Ghirlanda+13})
and it has been detected in X--ray flares (\citealt{Peng+14}).
Furthermore, \cite{Starling+12} and \cite{SparreStarling12} have presented systematic research of thermal signatures in X--ray emission.
According to the classification 
%made in 
of \cite{Ghirlanda+13}, GRB 151027A belongs to  \textit{Class III} of the thermal bursts because 
the thermal and non-thermal components coexist.
Fig. \ref{fig:lum_flare} shows the simultaneous evolution of the 0.3--1000 keV and 0.3--10 
keV luminosity of the two components. 

%The CPL component peak anticipates the thermal luminosity peak
% These two components simultaneously evolve in normalization within time (Fig. \ref{fig:parflare}),  with the CPL component reaching its maximum 
% before the BB component that instead 
 %arises from the CPL during its decreasing phase.  
%reaches its maximum when the CPL component is decreasing. 
%However, 
The X--ray flare of GRB\,151027A has a very luminous thermal component
($\sim$10$^{50}$ erg s$^{-1}$ near the peak) characterized by a low temperature 
($kT_{\rm BB} \sim 1$ keV, a factor of $\sim$10 lower than the typical temperature observed in GRB prompt emission, e.g. \citealt{Ryde04}).
Furthermore, the thermal luminosity peaks later than the non-thermal component and, at its maximum, it 
contributes to most of the total luminosity in the 0.3--10 keV and to  35\% of the 0.3--1000 keV luminosity.
In addition, the thermal component is still detected in the XMM-{\it Newton} late time spectrum with a luminosity $\sim$5$\times 10^{44}$ erg s$^{-1}$, corresponding to $\sim$8\% of the 0.3--10 keV emission.
In the following, we  discuss the possible origin of this blackbody emission.

The hypothesis that the observed blackbody emission is due to a Ib/c SN shock breakout has to be excluded.
In fact, the typical X--ray luminosity of such emission is $\sim$10$^{45}$ erg s$^{-1}$
(see e.g. \citealt{MatznerMcKee99,Campana+06,Ghisellini+07}), which is much lower than 
the BB luminosity ($\sim$10$^{50}$ erg s$^{-1}$) observed at the peak of the flare in GRB 151027A.

\cite{Piro+14} proposed a model based on the emission of a hot plasma cocoon (based on \citealt{Peer+06}) 
to explain the long-lasting thermal emission observed in the ultra-long GRB 130925A.  \cite{Starling+12} also used the cocoon expansion to explain the presence of thermal emission in X--ray spectra of GRB associated with a SN explosion.
Even this model cannot be applied to our case because the peak luminosity reached by the thermal 
component during the flare is larger than the expected value (which is of the order of $10^{47}$ erg s$^{-1}$ or greater).

Thermal emission is naturally predicted within the standard fireball scenario, when the relativistically 
expanding fireball releases the internal photons at the transparency radius (e.g. \citealt{Goodman86}; 
\citealt{Paczynski86}; \citealt{DM02}). 
Owing to the initial huge opacity of the fireball (optical depth $\tau \gg 1$), photons can reach 
the thermodynamic equilibrium and are characterized by a BB spectrum. 
  
\begin{figure}
\centering
\includegraphics[width=\columnwidth]{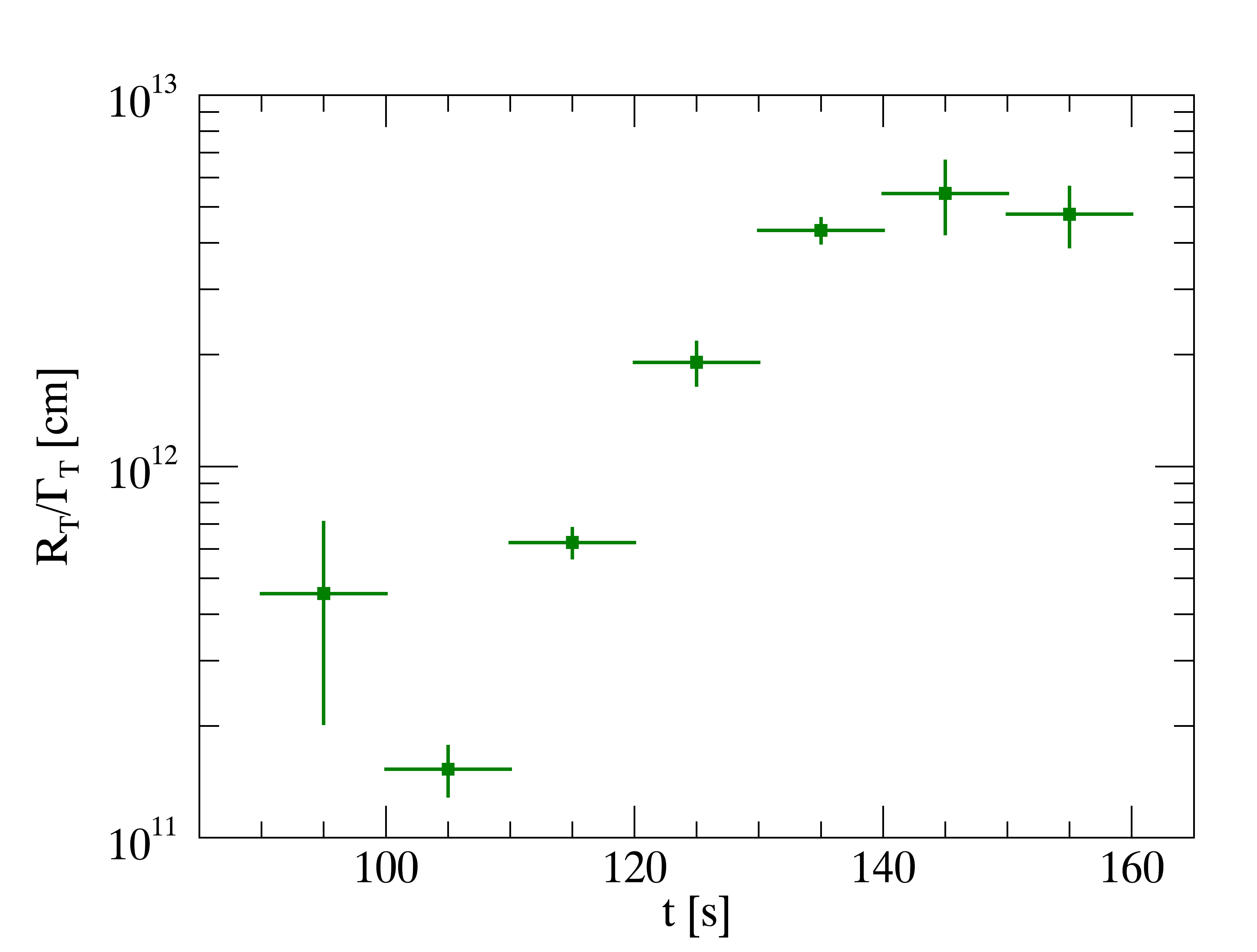}
\caption{$R_{\rm T}/\Gamma_{\rm T}$ as a function of time  obtained by the fit of time resolved spectra of the flare (Eq. \ref{eq:rsug}). 
}
\label{fig:RsuGflare}%
\end{figure}

Using the observables associated with  the BB spectrum, i.e. the temperature $kT_{\rm BB}$ and the flux 
$F_{\rm BB}$,  we can estimate the fundamental parameters of the fireball (see \citealt{Ghirlanda+13}). 
We can first obtain the ratio between the radius of the fireball and its bulk Lorentz factor 
$R_{\rm T}/\Gamma _{\rm T}$ when it becomes transparent: 

\begin{equation}
\frac{R_{\rm T}}{\Gamma_{\rm T}} = 
2.406 \frac{d_{\rm L}(z)}{(1+z)^2}\left(\frac{F_{\rm BB}}{\sigma T_{\rm BB}^4}\right)^{1/2} {\rm cm}
\label{eq:rsug}
\end{equation}
The evolution of this ratio during the third emission peak is shown in Fig. \ref{fig:RsuGflare}.
    
%Using the observables associated to the BB spectrum, i.e. the temperature $kT_{\rm BB}$ and the flux $F_{\rm BB}$,  we can estimate the fundamental parameters of the fireball. We can directly obtain the ratio between the radius of the fireball and its bulk Lorentz factor $R_{\rm T}/\Gamma _{\rm T}$ when it becomes transparent. 
%The evolution of this ratio during the third emission peak is shown in Fig. \ref{fig:RsuGflare}.

%To estimate these parameters, we are using the calculations developed in \cite{Ghirlanda+13} for the analysis of the thermal emission in GRB\,100507.

In order to test this hypothesis further, we need to make an assumption about when  transparency occurs: 

(i) It might happen during the acceleration phase when, owing to the high internal pressure, the fireball is still 
accelerating, converting its internal energy to bulk motion energy.
%In this regime 
%the $\Gamma$ grows up proportionally to $R$. 
%($\Gamma \propto R$). 
In this case, it is possible to estimate the distance from the central engine $R_0$, where the fireball is 
created, assuming an initial bulk Lorentz factor $\Gamma_{\rm i}=1$. We obtain $R_0 \sim 10^{11-12}$ cm.

%This relation allows us to estimate the distance from the inner engine $R_{0}$ where the fireball is created. This is possible because in the standard scenario the fireball is assumed to be created at $R_{0}$ with $\Gamma_{0} = 1$.

%Using the values of temperatures and fluxes obtained fitting the spectra and knowing the redshift of GRB\,151027A ($z = 0.81$) we find $R_{0} \sim 10^{11-12}$ cm.

\begin{figure}
\centering
\includegraphics[width=\columnwidth]{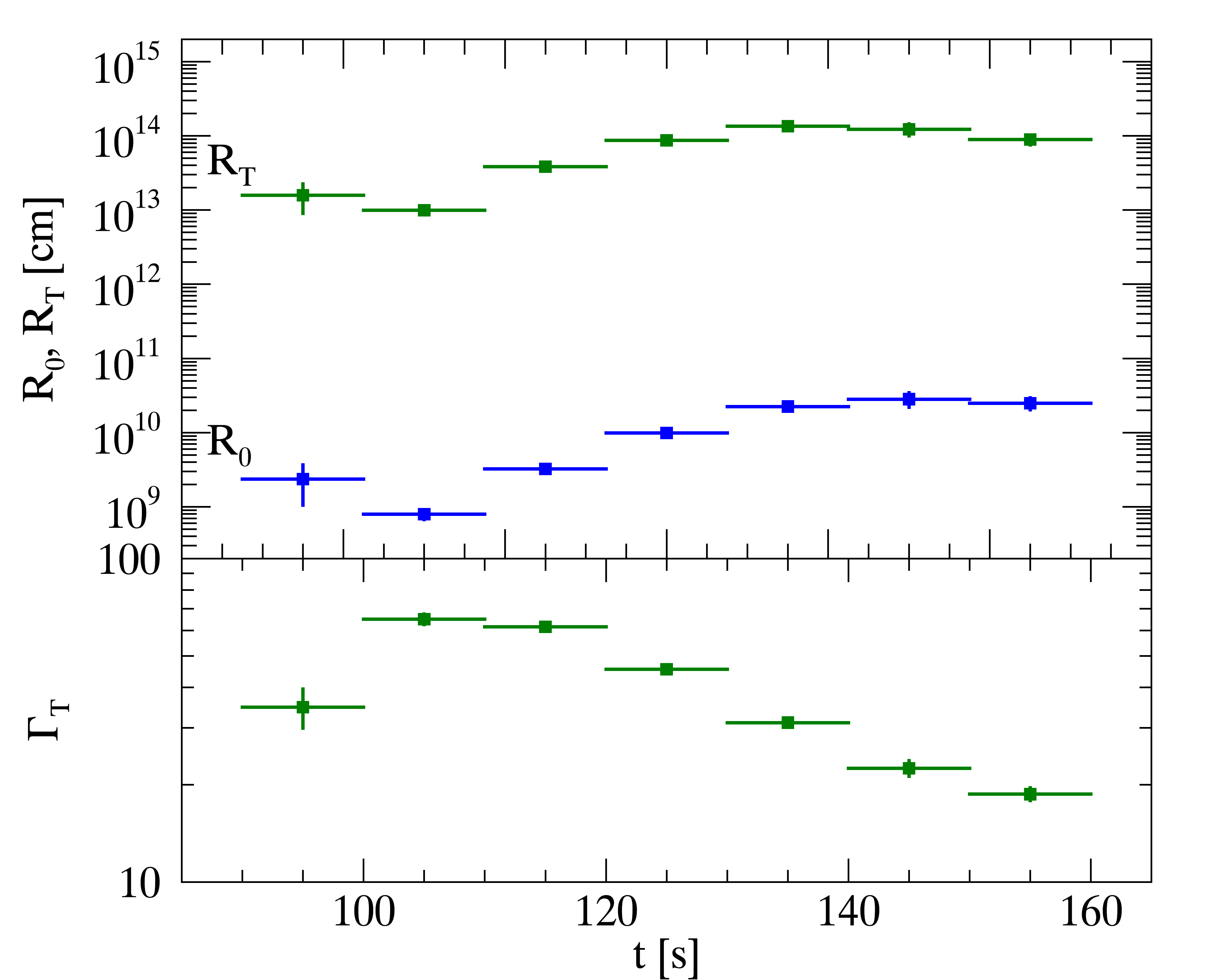}
\caption{$R_{\rm T}$, $\Gamma_{\rm T}$, and $R_0$ as functions of time in the hypothesis that the fireball 
becomes transparent during the coasting phase. 
In this case we used $\eta_{\rm BB}=10^{-2}$.
}
\label{fig:coast_flare}%
\end{figure}

(ii) It might happen during the coasting phase. 
The internal pressure is no longer sufficient to accelerate the fireball that proceeds with constant bulk Lorentz factor.
In this case combining Eq. \ref{eq:rsug} with the relations shown in \cite{DM02} we can 
obtain $R_{\rm T}$, $\Gamma_{\rm T}$, and $R_0$. 
%On the contrary with respect to 
Differently from the previous case, these values are not unequivocally determined because they depend on the 
blackbody radiative efficiency $\eta_{\rm BB}$\footnote{$\eta_{\rm BB}$ is the ratio between the energy emitted 
by the blackbody and the fireball total energy. 
It is smaller than the efficiency $\eta$ used in the modelling of GRB afterglows (typically $\eta\sim 0.2$,
cf. Eq. \ref{eq:eta}), since only a fraction of the emitted radiation is associated with the thermal component. }.
%This parameter represents the fraction of total energy released as radiation and it is typically estimated to be $\sim$ 20\% from the modelling of GRB afterglow (see e.g. \citealt{PK00} and \citealt{Ghisellini+09}).
%However, the right value we have to consider for $\eta$ is smaller because only a fraction of the emitted radiation is associated with the thermal component.
As in \cite{Ghirlanda+13} we use a radiative efficiency related to the thermal component of about 
$\eta_{\rm BB} \sim 10^{-2}$, since the blackbody flux varies from $\sim 5\%$ up to $\sim 50$\% 
of the non-thermal flux. 
Then, we find $R_{\rm T} = 10^{13-14}$ cm, $\Gamma_{\rm T} \sim 60$, and $R_{0} \sim 10^{9-10}$ cm (Fig. \ref{fig:coast_flare}).

In both cases the value of $R_0$ is much higher than expected. 
Assuming that the progenitor of long GRBs is a newly born %black hole
 compact object (a black hole or a magnetar)
produced by the core 
collapse of a Wolf--Rayet star (\citealt{Usov92, Duncan+92}; \citealt{Woosley93}; \citealt{McFWoos99}), we can suppose 
that the fireball should be formed near the central object%
%black hole
, at a few gravitational radii. 
For a compact object of $5-10\, M_{\odot}$, 
the gravitational radius is $\sim 10^6$ cm, so we expect that 
$R_0 \sim 10^7$ cm, a value much smaller than the obtained one.

%For standard values of efficiency $\eta$ the fireball should form even out of the progenitor star radius ($R_{*} \sim 10^{10}$ cm for a Wolf--Rayet star).

%Si forma molto lontano addirittura fuori dalla stella

%The radius $R_{0}$ obtained in the case of transparency during the acceleration can not be changed, but the one obtained during the coasting phase could be reduced only requiring an even lower radiative efficiency.
In case (ii), the value we should use for $\eta_{\rm BB}$ in order to get $R_{0} \sim 10^7$ 
is $\sim 10^{-4}$. Such a low radiative efficiency would imply an enormous burst of kinetic energy. 
Therefore, we  expect a very energetic afterglow that is in contrast with what we observe.

Another possible explanation of the significant thermal emission of GRB\,151027A is given by the 
``reborn fireball'' model (\citealt{GhiselliniReborn}). 
In this scenario the thermal emission is produced by plasma heated in the collision between the relativistic 
ejecta and the surrounding material released by the progenitor star during its final evolution stages.

If the optical depth after collision is large, a re-acceleration to relativistic speed  due to the dissipated internal energy can take place.
%If the collided material has a large optical depth at the impact, the dissipated energy 
%is held in the ``merged'' shell that is re--accelerated to relativistic velocities.
This process allows the creation of a reborn fireball with a larger initial radius
$R_0 \sim 10^{11}$ cm consistent with the large values inferred for GRB 151027A. 

\cite{GhiselliniReborn} 
assume the target material to be at rest with respect to the central engine. Nevertheless, in our case the relativistic shells that produced the first two prompt emission peaks should have interacted with such material first. For this reason, we must conclude that the optically thick target material was not there when the first prompt photons were emitted.

%assume that the collided material should be at rest with respect to the central engine frame. 
%Nevertheless, in this case the relativistic shells related to the first two prompt emission peaks were 
%supposed to interact with the dense external medium surrounding the progenitor. 
%For this reason we should exclude that the optically thick material was there when the first prompt 
%photons were emitted.
%gets trapped and the merged plasma  
%is re--accelerated to relativistic velocities.     

%Therefore, we assume that the GRB central engine is  responsible for the production of the optically thick material.
A possible way around this is to assume the GRB central engine itself is responsible for the production of the target material.
At the beginning%the black hole ejects the first shells that produce the initial part of the prompt emission. 
, shells that produce the initial part of the prompt emission are ejected.
Then a  denser and  slower shell is ejected, which  does not emit radiation since it is optically thick. 
After a quiescent period a quicker shell is ejected and it reaches the slower one.
In this scenario the reborn fireball is actually like an internal shock between a thick, 
mildly relativistic, massive shell with a quicker shell. 
The collision dissipates energy with non-negligible efficiency since the relative Lorentz factor can be large. 
The photons produced cannot escape because of the large opacity and the internal thermal energy can be used to 
re-accelerate the shell.
Beyond the photospheric radius the shell emits the blackbody radiation produced by the reprocessing 
of the trapped photons and a non-thermal component.
The decreasing emission of the flare is then due to the quenching of the radiation of the shell and to the off latitude emission.

\subsection{Modelling the afterglow}

In this section we propose a model for the afterglow light curve %in several bands 
from the XRT 0.3--10 keV flux to the optical $R$ band and 
to the radio frequencies. 
As was said before, both the X--ray and optical early time flux 
(for $t_{\rm obs} \, \lsim \, 500$ s) %cannot be considered 
%pure afterglow emission, because it 
is contaminated by the emission 
of the flare.
For this reason, we focus on the observed light curves
only for $t_{\rm obs} \, \gsim \, 500$ s.
At this epoch, the X-ray light curve shows the presence of a plateau phase (\citealt{Nousek+06}), which is usually related to a late time central engine activity (see e.g. \citealt{Zhang+06, DaiLu98, ZhangMeszaros01, Kumar+08, CorsiMesz09, Metzger+11, Leventis+14, vanEerten14, DuffMcF14}). However, we first attempt to model the multiwavelength long-lasting  emission of GRB 151027A as produced uniquely by the forward shock.

\subsubsection{ Model}

The modelling of the observed afterglow light curves 
has been performed with a semi-analytic model that combines 
the forward shock dynamics developed in \cite{Nava+13}
with the computation of the spectrum of the emitted radiation,
based on \cite{Nappo+14}, already used in \cite{Melandri+15}.
The model will be presented in more detail in a future paper
in preparation. Here we introduce only the most relevant features.

We assume that the blastwave starts moving at relativistic velocity with an initial bulk Lorentz factor
$\Gamma_0$, and with an initial kinetic energy $E_{\rm kin,0}$ that is linked to the emitted 
$\gamma$--ray isotropic energy $E_{\rm \gamma, iso}$ and the efficiency $\eta$ by 
\begin{equation}
E_{\rm kin,0} = (1-\eta)E_{\rm \gamma, iso}/\eta
\label{eq:eta}
\end{equation}
Then the fireball decelerates because of the interaction with the external medium 
and dissipates its energy (see \citealt{Nava+13} for an exhaustive treatment). 
We assume that %at each timestep 
a fraction $\epsilon_{\rm e}$ and $\epsilon_{\rm B}$ of 
the dissipated energy is distributed to the leptons and the magnetic field, respectively. 
The remaining energy is given to protons.
The energy is given to electrons %is used to inject them
with an energy distribution $Q(\gamma) \propto \gamma^{-p}$.
The leptons can cool for synchrotron and synchrotron self-Compton 
(SSC) %radiative emission 
emitting a fraction $\epsilon_{\rm rad}$ of their total energy.
%The emitted spectrum is normalised in each timestep to the bolometric
%luminosity obtained by the dynamical model.
At each time step we compute the following:
\begin{itemize}
\item the synchrotron spectrum  in the optically thin and in the self-absorbed regime;
\item the $Y$ Comptonization parameter;
\item the SSC spectrum;
\item the fraction $\epsilon_{\rm rad}$ of injected energy that is actually radiated.
\end{itemize}
The resulting spectrum is normalized at each time step to the bolometric luminosity obtained 
by the dynamical evolution.
The fireball is assumed spherical, but it is possible to insert a 
%an achromatic 
jet break
in the light curves when the beaming cone of width $1/\Gamma$ becomes larger than the jet opening angle 
$\theta_{\rm jet}$, which produces an achromatic steepening of the temporal index $\alpha$. %, used as opening angle of the jet.
We can describe the propagation of the forward external shock 
in a circumburst medium (CBM) with a generic density profile $n(R)$. 
In this work we will test only the two standard cases: 
homogeneous medium ($n(R)=$ const) and wind--medium ($n(R) \propto R^{-2}$); 
the first  describes the density profile 
%of a medium that 
%is similar to the normal 
typical of the
interstellar medium and the second  
describes the stratified density profile that can be produced 
by the intense stellar winds in the final stages of the Wolf--Rayet 
star evolution.

\begin{table}[h]
\centering
\begin{tabular}{ccc}
 \hline 
  Parameter &Homogeneous CBM &Wind CBM
\\
& ($s=0$) & ($s=2$)\\
\hline \hline
$\Gamma_0$ & 125 & 48 \\
$\epsilon_{\rm e}$ & 0.22 & 0.04 \\
$\epsilon_{\rm B}$ & $8 \times 10^{-5}$ & 0.06 \\
$n_0$ & 0.08 cm$^{-3}$ & $0.04\, A_{*}$ \\
%$1.2 \times 10^{34}$ cm$^{-1}$ \\
$\eta$ & 0.035 & 0.16 \\
$p$ & 2.4 & 2.65 \\
$\theta_{\rm jet}$ & 
$4.2^\circ\ $ & 
$6.3^\circ\ $ \\
$E_{\rm \gamma}$ [erg] & $1.1 \times 10^{50}$ & $2.4 \times 10^{50}$ \\
$E_{\rm \gamma,iso}$ [erg] & $3.98 \times 10^{52}$ & $3.98 \times 10^{52}$ \\
\hline
\multicolumn{3}{c}{Late Prompt Parameters} \\
\hline \hline
$T_{\rm A}\, [{\rm s}]$ & -- & $1\times 10^5$\\
$L_{\rm A}$ [erg/s]& -- & $1.3\times 10^{46}$\\
$\nu_{\rm b}\, [{\rm Hz}]$  & -- & $1\times 10^{16}$\\
$\beta_{\rm X}$ & -- & $1.0$\\
$\beta_{\rm o}$ & -- & $0.6$\\
$\alpha_{\rm fl}$ & -- & $0.4$\\
$\alpha_{\rm st}$ & -- & $2.0$\\
\hline \hline
\end{tabular}
\caption{\label{table:par} Parameters of GRB 151027A in the homogeneous and wind scenarios. 
The CBM density is expressed as $n(R)=n_0\times R^{-s}$, where $s=0$ represents the case of 
homogeneous CBM and $s=2$ the case of wind CBM, where the coefficient is expressed in unity 
of $A_*=3 \times 10^{35}$ cm$^{-1}$. 
%The limit on $\theta_{\rm jet}$ in the wind scenario is 
%obtained considering that no steepening in the light curve is observed until $t \sim  10^7$ s.
$L_{\rm A}$ is the 0.3--10 keV luminosity of late prompt at $T_{\rm A}$.
}
\end{table}

\subsubsection{ Homogeneous CBM scenario}
The best result obtained using this modelling hypothesis is represented
in the top panel of Fig. \ref{fig:lcaft}. 
The values used for the parameters in this scenario are given in the first column of Tab. \ref{table:par}.
The solution was obtained using standard values except for 
the efficiency $\eta$, which is about an order of magnitude smaller than typical values ($\eta \sim 0.2$).
In addition, a remarkably small value for $\epsilon_{\rm B}$ is used 
because, with a small magnetic field, the cooling frequency $\nu_{\rm c}$ is closer to the value inferred by the modelling of the $\sim$1.8$\times$10$^4$s SED with a pure synchrotron spectrum (see yellow region in panel 2 of Fig. \ref{fig:SedAft}).

The injection index of the electrons $p=2.4$ is consistent with the slope of the optical spectrum $\beta_{\rm o}$
measured at $t \sim 6\times 10^4$ s 
%($F_{\nu} \propto \nu^{-\beta}$, with 
($\beta_{\rm o}=(p-1)/2\simeq 0.7-0.9$).
Nevertheless, with this choice of $p$, 
%In the homogeneous density medium scenario 
a steepening of the light curve decay ($\Delta \alpha \sim 1$) for 
$t > 2 \times 10^5$ s is necessary to account for the optical and X--ray late time behaviour.
We interpret this achromatic steepening as the jet break (\citealt{Rhoads97}; \citealt{Sari+99}).
Using the standard relations for the jet break time in a homogeneous external medium (\citealt{Sari+99})
we can determine the jet opening angle $\theta_{\rm jet}$%:
%
%\begin{equation}
%\theta_{\rm jet} = 0.161 \left(\frac{t_{\rm j,d}}{1+z}\right)^{3/8}
%\left(\frac{n_0 \cdot \eta}{E_{\rm \gamma,iso,52}}\right)^{1/8}
%\label{eq:thjet_h}
%\end{equation}
%
%where $t_{\rm j,d}$ is the time of the jet break in days and $E_{\rm \gamma,iso,52}$ is 
%the isotropic equivalent energy of the prompt emission in unity of $10^{52}$ erg s$^{-1}$.
%We obtained an opening angle $\theta_{\rm jet} $
$= 4.2^\circ$. 

The collimation corrected 
$\gamma$-ray emission is $E_\gamma = E_{\rm \gamma,iso}(1-\cos \theta_{\rm jet})= 1.1 \times 10^{50}$ erg, 
where we used the isotropic energy obtained by {\it Fermi}--GBM integrated spectrum 
$E_{\rm iso}=3.98\times 10^{52}$ erg (cf. \S \ref{HEdata}).
This result has been compared with the 
$E_{\rm peak}-E_{\gamma}$ correlation (\citealt{Ghirlanda+04,Ghirlanda+07}) and the burst is $4\sigma$ off the best fitting line.
Using the {\it Fermi}--GBM rest frame $E_{\rm peak}=615$ keV, the jet opening angle 
that would make the GRB consistent with the $E_{\rm peak}-E_{\gamma}$ correlation 
%required to put the burst on the Ghirlanda correlation 
is $\theta_{\rm jet} = 14^\circ$, which should have generated 
an achromatic break in the light curves at $t \sim 5.8 \times 10^6$ s. No break is observed at this epoch.

Radio 5 GHz observations provide the main evidence that excludes the homogeneous model.
 Fig. \ref{fig:aft_radio} is a zoomed view of Fig. \ref{fig:lcaft} in which the late time 5 GHz model predictions with both homogeneous and wind model are compared with the data.
Indeed, the 5 GHz model in the homogeneous case  
is not compatible with the SRT upper limit at $3.5 \times 10^5$ s and with the EVN and VLBA observations. %(orange dashed line in Fig. \ref{fig:lcaft} and in Fig. \ref{fig:aft_radio}).
This significant incompatibility, in addition to the lack of strong evidence
of an achromatic break at $t\sim 2 \times 10^5$ s, leads us to conclude 
%that the jet break feature 
%required to reproduce the late time optical and X--ray emission is not reliable.
 % and the EVN observation of March 15.
%For all these reasons, we conclude 
that the homogeneous density profile does not provide a good 
%result for the 
modelling of the afterglow of GRB\,151027A\footnote{
The addition of a late prompt extra component (as described in \S \ref{sec:wind}) does not affect the conclusion since it could provide a better interpretation for the X-ray and the optical early emission, but it cannot influence the modelling of the late time radio band light curves. In particular, the flat evolution of 5 GHz flux density light curve (Fig. \ref{fig:aft_radio}) is not compatible with an external homogeneous medium.}.

%For what said in the preceding section, the behaviour of light curve at early times (for $t\, \lsim \, 500$ s) can not be explained with a standard afterglow forward shock scenario. For these reasons the aim is to model the X--ray, optical and radio afterglow light curves only for $t \, \gsim \, 500$ s.

\subsubsection{ Wind CBM scenario}\label{sec:wind}

The best solution obtained in the wind CBM scenario is plotted in the bottom panel of Fig. \ref{fig:lcaft}.
The corresponding parameters are shown in the second column of Table \ref{table:par}.
We adopted standard values for the parameters, except for the density parameter $n_0$, which is a 
factor of 25 smaller than the typical value %used in the typical wind profile 
$A_*=\dot{M_{\rm w}}/(4\pi m_{\rm p} v_{\rm w})=3\times 10^{35}$ cm$^{-1}$ 
obtained for a mass loss rate $\dot{M_{\rm w}}=10^{-5}$ M$_{\odot}$ yr$^{-1}$ and a 
wind velocity $v_{\rm w}=1000$ km s$^{-1}$, typical of a Wolf--Rayet star (\citealt{ChevalierLi99,ChevalierLi00}).

The injection index $p=2.65$ is compatible with the optical spectral slope of the SED at $\sim 6\times 10^4$ s and allows a description of the optical light curve temporal decay, which  is better than the value obtained in the homogeneous model.

The 22 GHz VLA observation at 0.78 days after the trigger (\citealt{Laskar18508}) deviates from the 
model prediction by a factor of $\sim 2.5$. This inconsistency can be explained with the scintillation 
caused by the circumburst medium (\citealt{Goodman97}), which  should modulate the early radio flux of GRB afterglows. 
For example, in the case of GRB 970508 (\citealt{Frail+97,Taylor+97}) the early time observed radio flux is strongly 
modulated up to a factor of $\sim 4$ at 8.46 GHz.

The prediction for the X--ray afterglow flux (blue dotted line in the bottom panel of Fig. \ref{fig:lcaft}) is much lower 
than the observed value. Furthermore, the X--ray  light curve profile shows some elements 
such as a plateau and a flare that cannot be explained in the standard forward shock scenario.
For these reasons, to model the X--ray emission we need to introduce another component 
of different origin.% from the forward shock afterglow.
%that describes the early time 
%shallow decay between $\sim 500$ s and $\sim 2000$ s and then the steeper decay until $\sim 10^5$ s.

The presence of this extra component is also suggested by the SEDs of the afterglow, especially the $\sim$1000 s SED obtained with UVOT and XRT (first panel of Fig. \ref{fig:SedAft}).
The X--ray flux is much higher than the extrapolation of the power law component of the UVOT emission and
requires another component to be consistent.
Instead,
the single spectral energy distributions taken at $\sim$1.8$\times$10$^4$ s and $\sim$6$\times$10$^4$ s (panels 2 and 3 in Fig. \ref{fig:SedAft}) %instead show profiles that 
are compatible with standard synchrotron emission spectrum
in slow cooling regime (i.e. the injection frequency is smaller than the cooling frequency, 
$\nu_{\rm i}<\nu_{\rm c}$; e.g. \citealt{PK00}) 
produced by a leptonic population with injection index $p\sim 2.4-2.8$. 
Nevertheless, the cooling break frequency is required to evolve as $\nu_{\rm c} \propto t^{-1.7}$.
Such evolution is incompatible with the standard scenario in both CBM density 
profiles\footnote{Following e.g. \citet{GranotSari02}, in the slow cooling regime, 
assuming an adiabatic evolution of the fireball, the cooling frequency evolves with 
time $\propto t^{-1/2}$ in the homogeneous case and $\propto t^{1/2}$ in the wind case. }
and can be accounted for assuming a further component in X--rays 
evolving differently from the standard forward shock evolution (\citealt{BM76,GranotSari02}) that we generically address as late prompt component (\citealt{GhiselliniLate}). 
% For these reasons we 
%interpret the observed light curve as the sum of two components: the standard forward shock afterglow emission 
%and a ``late prompt'' component % that describes the optical and X--ray early time shallow decay.
This %new component 
is 
%describe the observe
%introduce a ``late prompt'' component superimposed to the standard afterglow emission,
generated by a long-lasting central engine activity that ejects other shells that can move at relativistic
velocity, but with less energy and with a smaller Lorentz factor $\Gamma$. The physical mechanism that produces these shells relies on the nature of the central engine itself, is beyond the scope of the present work, and will not be discussed further.
%(as suggested by e.g. \citealt{GhiselliniLate}).
The modelling used for the late prompt component is taken from \cite{Ghisellini+09}, in which
\begin{itemize}
\item[-] the spectral shape is assumed to be constant in time and described by a broken power law:
\begin{equation}
L_{\rm late}(\nu , t) = L_{\nu_{\rm b}}(t)\begin{cases}
\left(\nu/\nu_{\rm b}\right)^{-\beta_{\rm o}} & \nu \leq \nu_{\rm b}\\
\left(\nu/\nu_{\rm b}\right)^{-\beta_{\rm X}} & \nu > \nu_{\rm b}\\
\end{cases}
\label{eq:LatePrSpe}
;\end{equation}
\item[-] the temporal evolution follows a smoothly broken power law profile:
\begin{equation}
L_{\rm late}(\nu , t) = L_{\rm late}(\nu , T_{\rm A})\frac{(t/T_{\rm A})^{-\alpha_{\rm fl}}}
{1+(t/T_{\rm A})^{\alpha_{\rm st}-\alpha_{\rm fl}}}
\label{eq:LatePrTime}
;\end{equation}
\item[-] the late prompt emission is present only in the optical and X--ray bands. Between the radio and the optical frequencies and beyond the X-ray frequencies there are exponential cutoffs. The cutoff frequencies are not considered as free parameters of the model.
\end{itemize}
There are seven parameters needed to describe the late prompt component: 
$\beta_{\rm X}$, $\beta_{\rm o}$, and $\nu_{\rm b}$ for the spectral behaviour; 
$\alpha_{\rm fl}$, $\alpha_{\rm st}$, and $T_{\rm A}$ for the temporal evolution; and  
$L_{\rm A}=\int L_{\rm late}(\nu , T_{\rm A})\, {\rm d}\nu$ for the normalization over the 0.3--10 keV band. 
The late prompt parameters adopted for the modelling are shown in Tab. \ref{table:par}.

\begin{figure}
\centering
\includegraphics[width=\columnwidth]{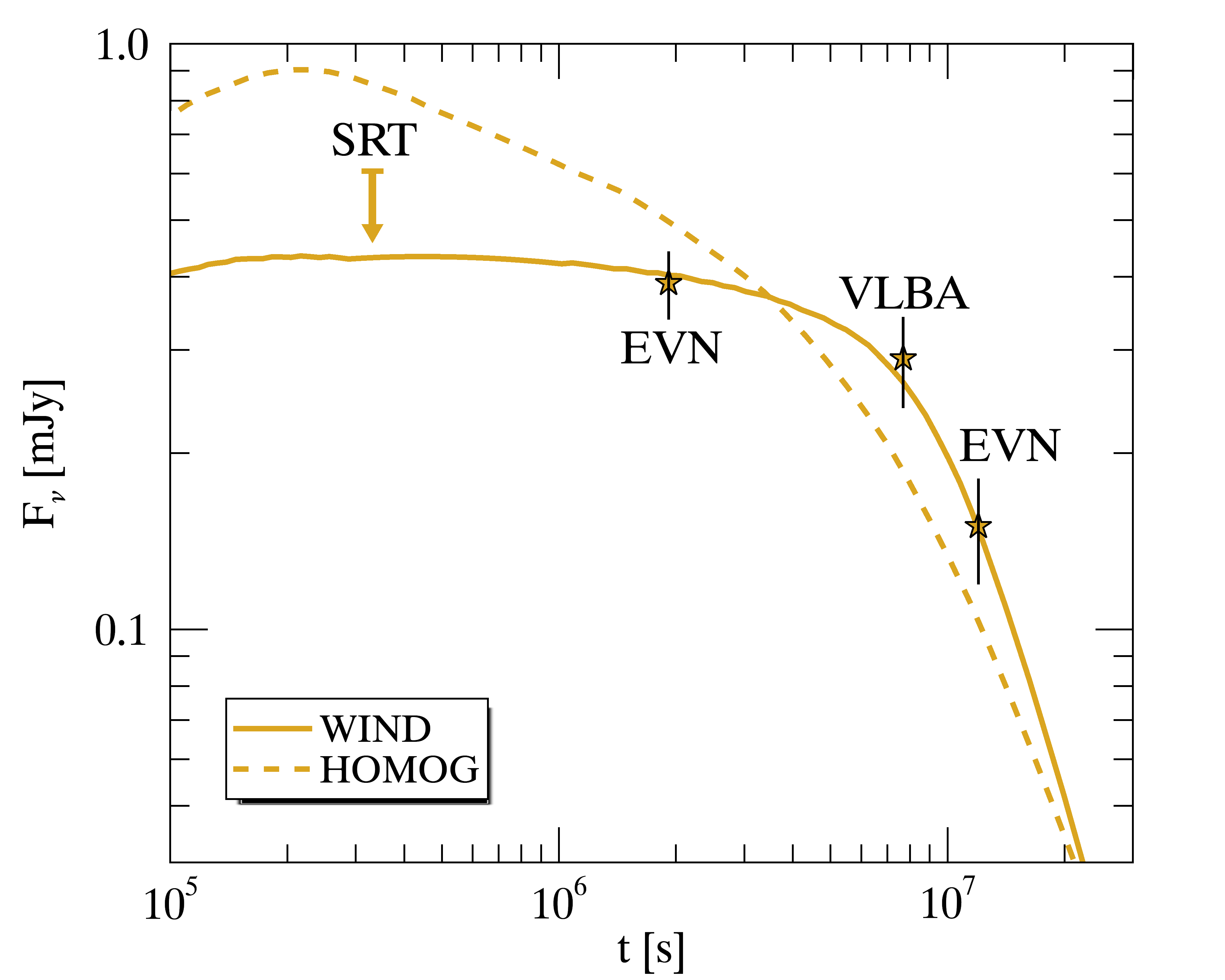}
\caption{Zoom of the 5 GHz band late time light curve (see Fig. \ref{fig:lcaft}). 
Dashed line represents the homogeneous model and solid line represents the wind model.
The 7 GHz SRT upper limit is compared with the 5 GHz model predictions.
}
\label{fig:aft_radio}%
\end{figure}

Late time EVN and VLBA 5 GHz radio observations and the SRT 7 GHz upper limit are in remarkable agreement with the solution of the wind density profile.
In particular the very flat evolution of the 5 GHz light curve (indicated by the SRT upper limit and the 5 GHz 
observation between $10^6$ s and $8\times 10^6$ s) can be explained with standard afterglow relations only in 
the case of a wind profile if the frequency of observation is between the self-absorption frequency 
$\nu_{\rm a}$ and the injection frequency $\nu_{\rm i}$.

The EVN observation of March 15 shows a very steep decrease in the 5 GHz flux that can be explained 
by the presence of a jet break between the observations of Feb 6 and Mar 15. 
Adopting a value of $t_{\rm j} = 8\times 10^6$ s, the value of the jet angle can be estimated with 
standard relations (see e.g. \citealt{ChevalierLi00}).
We obtain 
$\theta_{\rm j}= 6.3^\circ$, corresponding to a collimation corrected energy of 
$E_{\gamma}=2.4\times 10^{50}$ erg.
The late time jet break is consistent with the low value of the density parameter that is inferred from the modelling. In fact, in a low density environment the fireball takes a longer time to decelerate and  
thus the bulk
Lorentz factor $\Gamma$ becomes smaller than $\theta_{\rm jet}^{-1}$ at later times. 
In this case, the inferred value for $E_{\gamma}$ is fully consistent at 1.4$\sigma$ with the 
$E_{\rm peak}-E_{\gamma}$ correlation.\\

The compatibility with the $E_{\rm peak}-E_{\gamma}$ correlation can be considered as indirect evidence, in addition to the radio late time observations (Fig. \ref{fig:aft_radio}), that lead us to conclude that the 
blastwave of GRB\,151027A is expanding in a medium shaped by the wind of the stellar progenitor.
%wind CBM profile 
%best describes the behaviour of GRB\,151027A.
%-Descrizione parametri\\
%-Late prompt/ spettri dell'afterglow? Commenti...\\
%-Ottico e flare? Come sopra...
%Per cosa optiamo, perchÃ©?

\subsubsection{Possible evidence of SN?}

\begin{figure}
\centering
\includegraphics[width=\columnwidth]{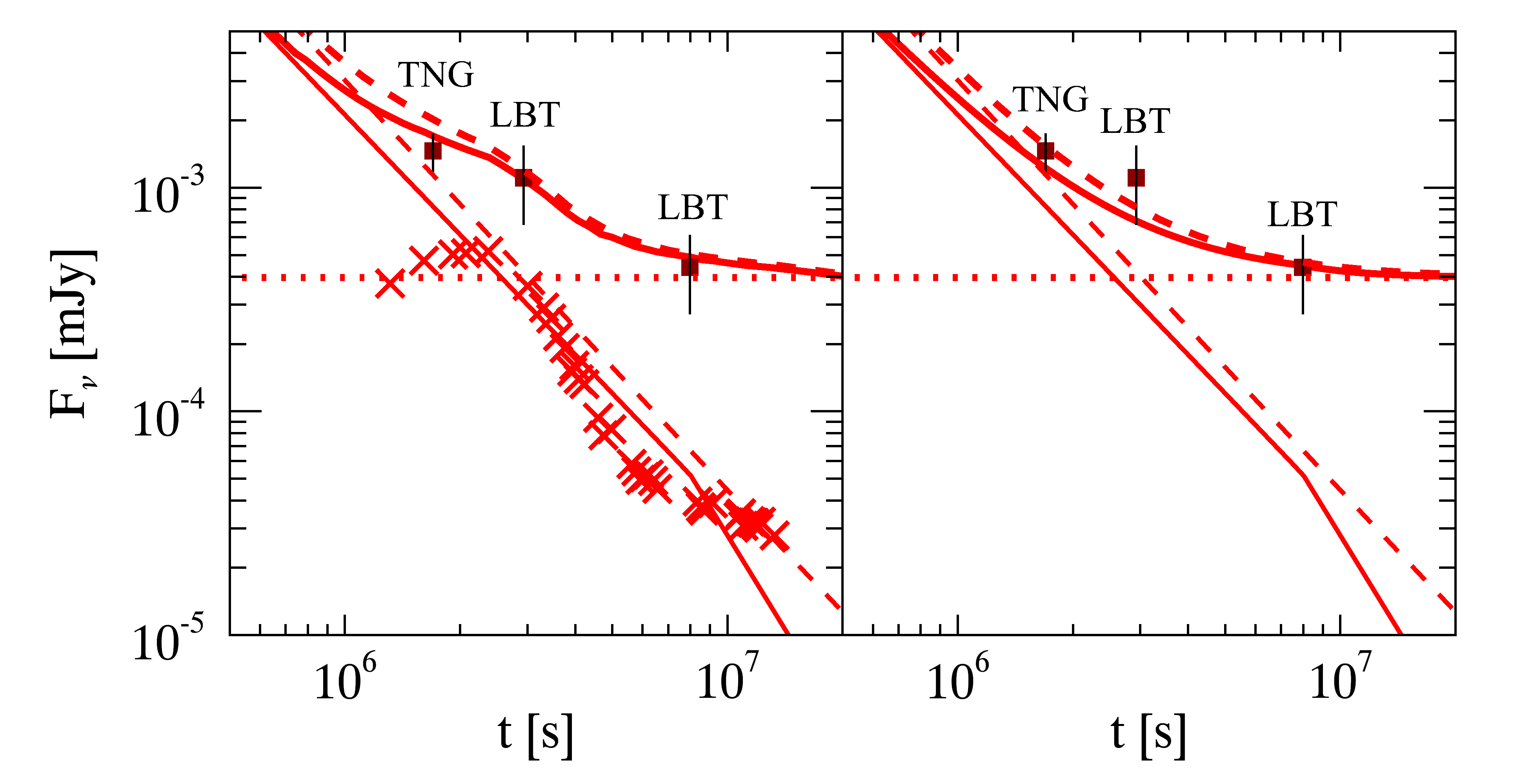}
\caption{Zoom of the $R$-band late time light curve (see Fig. \ref{fig:lcaft}). 
On the left data are modelled with afterglow (solid lines for wind medium, dashed lines for homogeneous medium), host galaxy (dotted line), and supernova component (red crosses), 
on the right only with afterglow and host galaxy. 
}
\label{fig:lcSN}%
\end{figure}

Late time optical observations after 19 days show a flattening in the light curve that can be explained 
by the presence of a supernova  and the host galaxy emission.
At 33 days a bump is identified in the optical light curve and it has been compared with a template of 
SN emission, namely the light curve of SN1998bw (\citealt{Galama+98}; \citealt{Clocchiatti+11}) rescaled at $z=0.81$ 
(red crosses in Fig. \ref{fig:lcaft}).
In particular, we synthesized the observed-frame $R$-band light curve of SN1998bw as it would appear if it occurred at that redshift using its rest-frame light curves and interpolating over frequency (\citealt{Cano13,Melandri+14}). We then included the SN contribution in the GRB late time light curve without applying any stretch (in flux or time).
In Fig. \ref{fig:lcSN} the late time $R$-band light curve already shown in Fig. \ref{fig:lcaft}  is zoomed
and compared with the modelling without the supernova contribution. 
Although the model without the supernova component is not incompatible with the LBT observation 
at $2.8\times 10^6$ s, the presence of a supernova emission leads to a better agreement with the late time observations.
If confirmed, it would be 
the eighth
most distant GRB/SN association ever discovered (see e.g. \citealt{HjorthBloom}; \citealt{Cano+16}).
The last LBT observation at 92 days after the trigger suggests the presence of an additional component 
that we interpret as the emission of the host galaxy. 
In this case the estimated flux density of the host galaxy is $\sim$ 0.4 $\mu$Jy in the $R$ optical band ($R_{\rm AB}\sim 25$), 
%in agreement
%with 
%a galaxy with a rest frame B-band absolute magnitude of $\sim$-18 and slope -1, 
similar to the flux density of
other GRB host galaxies at the same redshift (e.g. \citealt{Savaglio+09}; \citealt{Hjorth+12,Vergani+15,Perley+16}).
Only further observations (at least as deep as the LBT ones) can give a possible confirmation of this hypothesis.

%-SN 98bw riportata al giusto z.
%Possibile evidnenza bump latetime hce noi interpretiamo come... 
%Host galaxy di flux ... Possibile conferma di questa ipotesi si puÃ² avere solo con ulteriori osservaizoni as deep as LBT.

\section{Conclusions}

GRB 151027A, the 999th GRB observed by \textit{Swift}, is the first GRB with a bright flare 
starting $\sim$100 s after the GRB trigger and lasting $\sim$70 s, that has been simultaneously 
observed from the optical band up to the MeV $\gamma$-ray energy.
The time resolved spectral analysis of this flare indicates the presence of a blackbody component 
that provides up to  35\% of the total luminosity in the 0.3--1000 keV band.

In this work we discussed the possible origin of this thermal radiation.
%, studying in particular the case of blackbody emission
Since the radius and the luminosity of the blackbody emission were too large to be 
%explained with 
interpreted as the photospheric emission of a standard fireball model, we explored a reborn fireball scenario 
(\citealt{GhiselliniReborn}) in which the thermal radiation is produced by the energy dissipation due 
to the collision of a relativistic shell with a more massive, optically thick, slower one.

Intensive follow up campaigns provided a well-sampled multiwavelength afterglow  
light curve from X-rays to the radio band.

We interpreted the afterglow emission, where possible, in the standard forward shock scenario. 
We tested two CBM density profiles: the homogeneous (constant density, typical 
of the interstellar medium) and wind profile (with $n(R) \propto R^{-2}$, typical of the medium 
surrounding a massive star in the final stages of its evolution).
Since the X--ray light curve showed a plateau that cannot be explained by a standard afterglow 
behaviour, we needed to add a late prompt component (\citealt{GhiselliniLate}).

Late time radio observations provide  direct evidence of the better agreement of the data with the wind 
density profile model. 
In this case a jet break is observed, corresponding to a jet angle $\theta_{\rm jet}=6.3^\circ$.

Late time optical observations highlighted the presence of a bump in the light curve that can be 
interpreted as a supernova signature.
The late flattening of the $R$-band light curve allowed us to estimate the host galaxy flux $\sim$0.4 $\mu$Jy ($R_{\rm AB}\sim 25$).
%This GRB is an optimal candidate to test the standard external shock scenario with different density profiles
%and presents an interesting and well sampled flare showing a puzzling thermal emission.

\begin{acknowledgements}
The authors acknowledge the Italian Space Agency (ASI) for financial support through the ASI-INAF contract I/004/11/1. 
AR acknowledges support from Premiale LBT 2013.
This work made use of observations obtained with the Italian 3.6m Telescopio Nazionale Galileo (TNG) operated on the island of La Palma by the Fundaci\'on Galileo Galilei of the INAF 
(Istituto Nazionale di Astrofisica) at the Spanish Observatorio del Roque de los Muchachos of the 
Instituto de Astrofisica de Canarias under the program 
A32TAC\_5,
 and with the 8.4m Large Binocular Telescope (LBT) under the program 2015\_2016\_29. 
The LBT is an international collaboration among institutions in the Italy, United States, and Germany. LBT Corporation partners are: Istituto Nazionale di Astrofisica, Italy; The University of Arizona on behalf of the Arizona university system; LBT Beteiligungsgesellschaft, Germany, representing the Max-Planck Society, the Astrophysical Institute Potsdam, and Heidelberg University; The Ohio State University; and The Research Corporation, on behalf of The University of Notre Dame, University of Minnesota, and University of Virginia. 
We thank the TNG staff, in particular W. Boschin, and the LBT staff, in particular D. Paris, F. Cusano, and A. Fontana, for their valuable support with TNG and LBT observations.\\
The European VLBI Network is a joint facility of independent European, African, Asian, and North American radio astronomy institutes. Scientific results from data presented in this publication are derived from the following EVN project code: RG007.
The VLBA data used in the paper are obtained under the DDT Proposal VLBA/15B-382 (BG242).\\
SRT observations were performed in the framework of the
Astronomical Validation programme. 
The Sardinia Radio Telescope (SRT) is funded by the Department of
University and Research (MIUR), Italian Space Agency (ASI), and The
Autonomous Region of Sardinia (RAS) and is operated as National Facility
by the National Institute for Astrophysics (INAF).\\
This work made use of data supplied by the UK {\it Swift} Science Data Centre at the University of Leicester.\\
We acknowledge the referee for useful comments that
helped improve the manuscript.
\end{acknowledgements}


\begin{appendix}

%.
%XRT data were first automatically extracted from the {\it Swift}/XRT website (\url{http://www.swift.ac.uk/xrt_spectra/}) but we noticed an abnormal photon redistribution under 0.5 keV. Then, we manually extracted source spectra,background and ancillary response file for each time--bin and correcting for the pile-up effect using the xselect (mettere versione?altro?) tool. Before the analysis they were also rebinned 

%XRT source, background and ancillary response files for each time--bin were extracted from the {\it Swift}/XRT website (\url{http://www.swift.ac.uk/xrt_spectra/}). 
%Before the analysis they were also rebinned with the \textit{grppha} tool by requiring a minimum of 20 counts per bin. 

\section{Additional tables}\label{app:data}
\begin{table*}[]
\caption[]{Optical $R$--band light curve}
\centering
\begin{tabular}{cccc|cccc}
 \hline 
    
  $t-t_0$ [s]&
  $t_{\rm exp}$ [s]&
  $F_{\nu}$ [mJy]& Ref. &  
  $t-t_0$ [s]&
  $t_{\rm exp}$ [s]&
  $F_{\nu}$ [mJy]& Ref.
   \\
  \hline \hline
     24 & 12 & 11.08 $\pm$ 2.053 & [1] &	   $3.949\times 10^4$ & 150  & 0.7666   $\pm$ 0.04238 & [8] \\
   79.2 & 30 &  9.388 $\pm$ 1.739 & [2] &	 $4.195\times 10^4$ & 150  & 0.7187   $\pm$ 0.04637 & [8] \\
   126 & 30 & 12.84 $\pm$ 2.379 & [2] &	   $4.457\times 10^4$ & 150  & 0.6033   $\pm$ 0.04449 & [8] \\
   171.6 & 30 & 8.027 $\pm$ 1.487 & [2] &	   $4.520\times 10^4$ & 180  & 0.5967   $\pm$ 0.01649 & [9] \\
   217.2 & 30 & 11.08 $\pm$ 2.053 & [2] &	   $4.609\times 10^4$ & 180  & 0.6283   $\pm$ 0.01736 & [9] \\
   264 & 30 & 11.39 $\pm$ 2.11 & [2] &	   $4.822\times 10^4$ & 1200 & 0.5018  $\pm$  0.05558 & [10] \\
   1366 & 30 & 4.705 $\pm$ 0.8716 & [2] &	   $5.077\times 10^4$ & 1200 & 0.5018  $\pm$  0.04165 & [10] \\
   1468 & 30 & 3.737 $\pm$ 0.6923 & [2] &	   $5.242\times 10^4$ & 150 & 0.6376  $\pm$  0.03525 & [11] \\
   1501 & 30 & 6.033 $\pm$ 1.118 & [2] &	   $5.260\times 10^4$ & 150 & 0.4411 $\pm$ 0.03253 & [11] \\
   1556 & 30 & 5.207 $\pm$ 0.9645 & [2] &	   $5.499\times 10^4$ & 180  & 0.4186   $\pm$ 0.02237 & [9] \\
   1612 & 30 & 3.986 $\pm$ 0.7384 & [2] &	   $5.643\times 10^4$ & 180  & 0.4577   $\pm$ 0.01855 & [9] \\
   1823 & 30 & 4.06 $\pm$ 0.7522 & [2] &	   $5.740\times 10^4$ & 180  & 0.3942   $\pm$ 0.01707 & [9] \\
   1889 & 30 & 3.377 $\pm$ 0.6256 & [2] &	   $5.917\times 10^4$ & 180  & 0.3213   $\pm$ 0.02192 & [9] \\
   1973 & 30 & 4.098 $\pm$ 0.7591 & [2] &	   $6.344\times 10^4$ & 180  & 0.3109   $\pm$ 0.03385 & [9] \\
   2087 & 180 & 3.196 $\pm$ 0.592 & [3] &	   $6.451\times 10^4$ & 180  & 0.2334   $\pm$ 0.02149 & [12] \\
   2105 & 30 & 4.619 $\pm$ 0.8557 & [2] &	   $6.452\times 10^4$ & 180  & 0.3123   $\pm$ 0.01871 & [9] \\
   2152 & 30 & 4.927 $\pm$ 0.9127 & [2] &	   $7.776\times 10^4$ & 600 & 0.5978 $\pm$ 0.02203 & [13] \\
   2200 & 30 & 8.405 $\pm$ 1.557 & [2] &	   $1.119\times 10^5$ & 120  & 0.1108   $\pm$ 0.01022 & [6] \\
   2360 & 300 & 6.495 $\pm$ 1.203 & [3] &	   $1.134\times 10^5$ & 120  & 0.09291   $\pm$ 0.008557 & [14] \\
   2523 & 30 & 7.595 $\pm$ 1.407 & [2] &	   $1.318\times 10^5$ & 1800 & 0.1108  $\pm$  0.01022 & [15] \\
   2984 & 300 & 5.303 $\pm$ 0.9824 & [3] &	   $1.358\times 10^5$ & 900 & 0.09216 $\pm$ 0.0085 & [15] \\
   3250 & 300 & 5.207 $\pm$ 0.9645 & [3] &	   $1.476\times 10^5$ & 360 & 0.1602 $\pm$ 0.04482 & [16] \\
   5083 & 300 & 4.331 $\pm$ 0.8022 & [3] &	   $1.506\times 10^5$ & 3000 & 0.05018 $\pm$ 0.005558 & [17] \\
   8012 & 300 & 3.377 $\pm$ 0.6256 & [3] &	   $1.668\times 10^5$ & 300 & 0.04705 $\pm$ 0.001734 & [18] \\
   $1.045\times 10^4$ & 300 & 2.809 $\pm$ 0.5204 & [3] &	   $2.081\times 10^5$ & 3300 & 0.03877 $\pm$ 0.001786 & [19] \\
   $1.173\times 10^4$ & 300 & 2.358 $\pm$ 0.4368 & [3] &	   $2.523\times 10^5$ & 300 & 0.02151 $\pm$ 0.001586 & [18] \\
   $1.286\times 10^4$ & 300 & 1.805 $\pm$ 0.3344 & [3] &	   $2.915\times 10^5$ & 4680 & 0.02942 $\pm$ 0.001378 & [17] \\
   $1.893\times 10^4$ & 2220 & 1.89  $\pm$  0.08709 & [4] &	   $3.798\times 10^5$ & 4920 & (105.8 $\pm\, 6.827)\times 10^{-4}$ & [19] \\
   $2.092\times 10^4$ & 60  & 1.332 $\pm$ 0.1229 & [5] &	   $4.639\times 10^5$ & 2280 & (84.05 $\pm\, 9.309)\times 10^{-4}$ & [19] \\
   $2.662\times 10^4$ & 60  & 1.108 $\pm$ 0.1022 & [6] &	   $1.698\times 10^6$ & 2640 & (14.61 $\pm\, 2.706)\times 10^{-4}$ & [20] \\
   $2.700\times 10^4$ & 300  & 0.5862 $\pm$ 0.1080 & [7] &	   $2.929\times 10^6$ & 636 & (11.08 $\pm\, 4.175)\times 10^{-4}$ & [21] \\
   $3.089\times 10^4$ & 60  & 0.7187 $\pm$ 0.1331 & [2] &	   $7.977\times 10^6$ & 4812 &  $(4.411 \pm\, 1.662)\times 10^{-4}$ & [21] \\
\hline \hline
\end{tabular}

\tablebib{[1]~\citet{Wren18495}; [2] \citet{Pozanenko18635}; [3] \citet{Wiggins18539}
[4] \citet{Yano18491}; [5] \citet{Xu18485}; [6] \citet{Xin18515}; [7] \citet{Zhang18493}; [8] \citet{SahuAnupama18609};
[9] \citet{Oksanen18567}; [10] \citet{Hentunen18503}; [11] \citet{Sonbas18518}; [12] \citet{Cano18552}
[13] \citet{Dichiara18510}; [14] \citet{Zhang18513}; [15] \citet{Moskvitin18521}; [16] \citet{ProtKov18533};
[17] \citet{Kozlov18558}; [18] \citet{CenkoPerley18537}; [19] \citet{Mazaeva18559}; 
[20] This work: TNG observation; [21] This work: LBT observations
%(2) \citet{phillips87}; (3) \citet{barbon90}; (4) \citet{wells94};
%(5) \citet{mazzali93}; (6) \citet{gomez98}; (7) \citet{kirshner93};
%(8) \citet{patat96}; (9) \citet{salvo01}; (10) \citet{branch03};
%(11) \citet{jha99}.
}
\label{tabottico}
\end{table*}

\begin{table*}[]
\caption{Log of VLBI observations. 
\label{t.logvlbi} }
\center
\begin{tabular}{lllll}
\hline
Date & Frequency & On source time & Main duty cycle & Notes \\
 & (GHz) & (hrs) & (Cal-Tar-Cal) & \\
\hline \hline
2015 Nov 18 & 5.0 & 7.0   & 1-3.5-1   & EVN \\
2016 Jan 24 & 5.0 & 0.75  & 1-3.5-1   & VLBA, HN no fringes \\
2016 Jan 24 & 8.4 & 0.81  & 1-3.5-1   & VLBA, HN no fringes \\
2016 Jan 30 & 8.4 & 0.86  & 1-3.5-1   & VLBA, LA, MK, limited time \\
2016 Jan 30 & 15  & 1.00  & 1.5-3-1.5 & VLBA, LA, MK, limited time \\
2016 Feb 6  & 15  & 2.07  & 1.5-3-1.5 & VLBA, MK did not observe \\
2016 Mar 15 & 5.0 & 7.0 & 1-3.5-1 & EVN \\ 
\hline\hline
\end{tabular}
\tablefoot{HN: Hancock; LA: Los Alamos; MK: Mauna Kea. }
\end{table*}
%_____________________________________________________________

\begin{table*}
\caption{VLBI results. 
%{\bf MANCA OSSERVAZIONE EVN 15 MARZO}
\label{t.resvlbi} }
\centering
\begin{tabular}{llllllll}
\hline
$t_{\rm obs}$ & Array & $\nu$ & $\sigma_\nu$ & $B_\nu$ & $B_\nu/\sigma_\nu$ &HPBW & $S_\nu$ \\
 $[{\rm s}]$ & & [GHz] & [\mujb] & [\mujb] & & [mas $\times$ mas, $^\circ$] & [mJy] \\
\hline \hline
1.9 $\times$ 10$^6$ & EVN & 5.0 & 28 & 407 & 14.5 & $8.67 \times 6.57, \ -5.5$ & $0.39 \pm 0.05$ \\
7.7 $\times$ 10$^6$ & VLBA & 5.0 & 20 & 189 & 9.5 & $4.94 \times 1.17, \ -28.9$ & $0.29 \pm 0.05$\\
7.9 $\times$ 10$^6$ & VLBA & 8.4 & 37 & 194 & 5.2 & $1.44 \times 0.88, \ -59.7$ & $0.18 \pm 0.03$\\
8.5 $\times$ 10$^6$ & VLBA & 15 & 34 & 150 & 4.4 & $1.53 \times 0.73, \  40.6$ & $0.14 \pm 0.03$\\
1.2 $\times$ 10$^7$ & EVN & 5.0 & 22 & 124 & 5.6 & $6.0 \times 4.6, \ -40.3$ & $0.15 \pm 0.03$ \\
\hline \hline
\end{tabular}

\end{table*}

\end{appendix}
\end{document}